\documentclass[structabstract]{aa}  
\usepackage{natbib}
\usepackage{lscape}
\usepackage{rotating}
\usepackage[T1]{fontenc}
\usepackage[utf8]{inputenc} 
\usepackage[toc,page]{appendix}
\usepackage{graphicx}
\usepackage{enumerate}
\usepackage{amsmath}
\usepackage{subcaption}
\usepackage{txfonts}
\usepackage{float}
\usepackage{textcomp}
\usepackage{gensymb}
\usepackage[colorlinks=true,linkcolor=blue,citecolor=blue]{hyperref}
\usepackage{color}
\usepackage{mathtools}
\usepackage{ mathrsfs }
\usepackage[para,online,flushleft]{threeparttable}

\begin{document}

  \title{The origin of pulsating ultra-luminous X-ray sources:\\  Low- and intermediate-mass X-ray binaries containing neutron star accretors}

   \author{D. Misra
          \inst{1}\fnmsep\thanks{e-mail: devina.misra@unige.ch},
          T. Fragos\inst{1},
          T.~M. Tauris \inst{3,4},
          E. Zapartas\inst{1},
          D.~R. Aguilera-Dena\inst{2,5}
          }

   \authorrunning{Misra et al.}
    \titlerunning{LMXBs/IMXBs as NS ULX Sources}

   \institute{Observatoire de Gen\`eve, Universit\'e de Gen\`eve, 24 rue du G\'en\'eral-Dufour, 1211 Gen\`eve 4, Switzerland
         \and
              Argelander Institut f\"ur Astronomie, Universit\"at Bonn,
              Auf dem H\"ugel 71, 53121 Bonn, Germany 
        \and
              Aarhus Institute of Advanced Studies (AIAS), Aarhus University, H{\o}egh-Guldbergs~Gade~6B, 8000~Aarhus~C, Denmark
        \and
              Department of Physics and Astronomy, Aarhus University, Ny Munkegade 120, 8000~Aarhus~C, Denmark
        \and
              Max-Planck-Institut f\"ur Radioastronomie, Auf dem H\"ugel 69, 53121 Bonn, Germany
             }

   \date{Accepted on July 20, 2020}

% \abstract{}{}{}{}{} 
% 5 {} token are mandatory
 
  \abstract
  % context heading (optional)
  % {} leave it empty if necessary  
    {Ultra-luminous X-ray sources (ULXs) are those X-ray sources located away from the centre of their host galaxy with luminosities exceeding the Eddington limit of a stellar-mass black hole ($L_X>10^{39}\;{\rm erg\,s}^{-1}$). Observed X-ray variability suggests that ULXs are X-ray binary systems. The discovery of X-ray pulsations in some of these objects (e.g. M82~X-2) suggests that a certain fraction of the ULX population may have a neutron star as the accretor.} 
  % aims heading (mandatory)
   {We present systematic modelling of low- and intermediate-mass X-ray binaries (LMXBs and IMXBs; donor-star mass range $0.92$--$8.0$~M$_{\odot}$ and neutron-star accretors) to explain the formation of this sub-population of ULXs.}
  % methods heading (mandatory)
   {Using MESA, we explored the allowed initial parameter space of binary systems consisting of a neutron star and a low- or intermediate-mass donor star that could explain the observed properties of ULXs. These donors are transferring mass at super-Eddington rates while the accretion is limited locally in the accretion disc by the Eddington limit. Thus, our simulations take into account beaming effects and also include stellar rotation, tides, general angular momentum losses, and a detailed and self-consistent calculation of the mass-transfer rate.}
  % results heading (mandatory)
   {Exploring the initial parameters that lead to the formation of neutron-star ULXs, we study the conditions that lead to dynamical stability of these systems, which depends strongly on the response of the donor star to mass loss. Using two values for the initial neutron star mass ($1.3$~M$_{\odot}$ and $2.0$~M$_{\odot}$), we present two sets of mass-transfer calculation grids for comparison with observations of NS ULXs. We find that LMXBs/IMXBs can produce NS-ULXs with typical time-averaged isotropic-equivalent X-ray luminosities of between $10^{39}$ and $10^{41}\;{\rm erg\,s}^{-1}$ on a timescale of up to $\sim\!1.0\;{\rm Myr}$ for the lower luminosities. Finally, we estimate their likelihood of detection, the types of white-dwarf remnants left behind by the donors, and the total amount of mass accreted by the neutron stars. }
  % conclusions heading (optional), leave it empty if necessary 
   {We show that observed super-Eddington luminosities can be achieved in LMXBs/IMXBs undergoing non-conservative mass transfer while  assuming geometrical beaming. We also compare our results to the observed pulsating ULXs and infer their initial parameters. Our results suggest that a large subset of the observed pulsating ULX population can be explained by LMXBs/IMXBs in a super-Eddington mass-transfer phase.}


   \maketitle

\section{Introduction}\label{sec:intro}

Ultra-luminous X-ray sources (or ULXs) are extra-galactic sources first discovered by \cite{1989ARA&A..27...87F} that have been observed to have X-ray luminosities of $L_\mathrm{X}\gtrsim10^{39}$~erg~s$^{-1}$.  X-ray variability  suggests that ULXs are binaries with a non-degenerate star, referred to as a donor, transferring mass onto a compact object, the accretor \citep{1976MNRAS.175..395B, 2009MNRAS.397.1061H}. These objects often dominate the total X-ray emission of their host galaxy. They are generally too bright to be low-mass X-ray binaries (LMXBs; for which $L_\mathrm{X}\lesssim10^{37}$~erg~s$^{-1}$), and too dim when compared to active galactic nuclei (AGNs; for which $L_\mathrm{X}\gtrsim10^{41}$~erg~s$^{-1}$) with their position being off-centre in their host galaxy \cite[see][for a review on the observational properties of ULXs]{2017ARA&A..55..303K}.

Eddington luminosity is the limit above which any accretion onto the compact object is stopped by outgoing radiation. In general, ULX luminosities far exceed the Eddington limit of stellar-mass black holes (which is of the order of $10^{39}$~erg~s$^{-1}$) provided the emission is assumed to be isotropic. Initial suggestions as to the nature of the accretor include intermediate-mass black holes (IMBHs; with masses $\gtrsim100$~M$_{\odot}$) accreting at sub-Eddington rates \citep{1999ApJ...519...89C,2001ApJ...562L..19E, 2006ASPC..352..121M, 2007Natur.445..183M}, or stellar-mass black holes (SMBHs) accreting at super-Eddington rates. For the latter, in order not to exceed the Eddington limit locally while still producing the observed luminosity, magnetic fields and geometric X-ray beaming effects have been proposed as an explanation \citep[e.g.][]{2001ApJ...552L.109K,2014arXiv1411.5434C}. 

Since the first gravitational wave event, GW150914, detected by LIGO \citep{2016PhRvD..93l2003A}, there has been vast renewed interest in studying the formation of double compact objects. Many of the proposed formation channels for these double compact objects predict that the binary will go through one or more phases of super-Eddington accretion onto a compact object.  \cite{2017A&A...604A..55M} explored how some observed ULXs might indeed be progenitors of coalescing double compact object binaries, specifically through the chemically homogeneous channel. In this type of binary evolution, the star would avoid large expansion of the envelope, and therefore the binary would avoid a common-envelope (CE) phase \citep{2016MNRAS.458.2634M}. \cite{2017MNRAS.472.3683F} compared observation trends in the number of ULXs per unit of star-formation rate and as a function of metallicity of the host galaxy to the merger rate of binary BHs. These latter authors found that the majority of ULXs could be progenitors of binary BH mergers. Therefore, studying the formation of ULXs could shed light on merging binary compact objects, and vice versa.

In recent decades, debate over the nature of ULXs has focused on whether they are SMBHs or IMBHs, and neutron stars (NSs) have not been considered. However, theoretical studies have extended ULX binary models to include NSs \citep[e.g.][]{2001ApJ...552L.109K, 2009MNRAS.393L..41K}. That is until \cite{2014Natur.514..202B} reported the first ever observations of X-ray pulsations in the M82 galaxy with a period of about 1.37~seconds and a 2.52-day sinusoidal modulation in the ULX~X-2. Detection of X-ray pulsations in ULXs suggests the presence of NSs as accretors instead of BHs, at least in a certain fraction of the ULX population where such pulses have been observed. This is because X-ray pulsations are characteristic of accreting NSs with radiation being emitted along their magnetic poles as they rotate about their axes. Therefore, the question of accreting NSs in binaries with relatively massive donors was raised. This discovery was followed by the detection of X-ray pulsations in several other ULXs \citep{2017MNRAS.466L..48I, 2017Sci...355..817I, 2018MNRAS.476L..45C,2018NatAs...2..312B, 2019MNRAS.tmpL.104S,2019ApJ...879...61Z,2019arXiv190604791R}. 

The two most prominent questions regarding the nature of NS ULXs refer to the emission mechanism and the formation channel.
For the first question, a number of mechanisms are invoked to explain super-Eddington luminosities. With the apparent extremely high mass-transfer rate, a lot of the transferred mass could be blown away by strong radiation outflows \citep{1973A&A....24..337S}. \cite{2006MNRAS.370..399B} applied this idea to SS433 where observations of massive outflows suggest that the source is a ULX seen from the side. The absorption features associated with these outflows have been observed in some but not all ULXs; see for example Holmberg IX X-1 and NGC 1313 X-1  \citep{2012MNRAS.426..473W}. However, this does not invalidate the theory, as beamed X-ray emission would not be visible unless the observer had a direct line of sight to the accreting compact object, down the collimating funnel, in which direction the outflows are limited. \cite{1973A&A....24..337S}, \cite{2002ApJ...568L..97B}, and \cite{2007MNRAS.377.1187P} explored the idea of the presence of strong optically thick outflows that blow away some part of the disc from where radiation can escape. \cite{1973A&A....24..337S} and \cite{2007MNRAS.377.1187P} also suggested the formation of  geometrically thick accretion discs. This structure would cause the emission to be beamed, and therefore the observed isotropic-equivalent luminosity would be much higher than the intrinsic one. \cite{2001ApJ...552L.109K} postulated that based on the assumption of mild beaming, intermediate- and high-mass X-ray binaries (IMXBs and HMXBs) undergoing mass transfer on a thermal timescale would be the best candidates for ULXs. The effect of beaming on the emission has, in general, been explored \citep{2009MNRAS.393L..41K,2017MNRAS.468L..59K,2019ApJ...875...53W}. In addition to that, NSs with strong magnetic fields (around $10^{14}$~G) reduce the electron-scattering cross-section and could anchor the infalling matter to accretion columns above the magnetic poles and thereby produce sufficiently high luminosities \citep{1976MNRAS.175..395B, 2018NatAs...2..312B}. \cite{2017ApJ...845L...9T} carried out general relativistic radiation magnetohydrodynamics simulations of super-Eddington accretion onto a non-rotating, magnetised NS and found a spin-up rate of $\sim -{10}^{-11}~{\rm{s}}~{{\rm{s}}}^{-1}$, which is consistent with observations. In contrast, \cite{2019MNRAS.485.3588K} suggested that the observed ULX properties are explained by NSs with normal magnetic fields and not by the presence of magnetars. For the specific case of M82~X-2, \cite{2014arXiv1410.8745L} suggested the presence of an optically thick accretion disc that acts as a curtain and shields some of the outgoing radiation, thus allowing for super-Eddington luminosities driven by Roche-lobe overflow (RLO). Finally, \cite{2011ApJ...739...42W} studied accretion discs in weakly magnetised NSs, finding that at super-Eddington rates, the magnetic field has little effect on the accretion disc. 

Despite a lot of research done in the field, the question surrounding how ULXs attain super-Eddington luminosities is still open and an active field of research. Regarding a possible formation channel for NS ULXs,  a lot of work has been carried out to address evolution and mass transfer in X-ray binaries. \cite{1999A&A...350..928T} performed non-conservative mass-transfer calculations of low-mass X-ray binaries with a $1.3$~M$_{\odot}$ NS and found that the binaries can undergo very high mass-transfer rates (super-Eddington by a factor of $\sim 10^4$) for donors with deep convective envelopes (mass range $1.6$--$2.0$~M$_{\odot}$). An important X-ray source in the context of high mass-transfer rates is Cygnus X-2, which is an X-ray binary containing a NS. Cygnus X-2 has a mass ratio of $\approx 0.34~(=M_{\mathrm{donor}}/M_{\mathrm{acc}})$, so for an estimated accretor mass of about $1.78$~M$_{\odot}$ the donor has a mass of $0.6$~M$_{\odot}$ \citep{1998ApJ...493L..39C, 1999MNRAS.305..132O}. \cite{1999MNRAS.309..253K} argued that the donor star lost a lot of mass ($\sim 3.0$~M$_{\odot}$) in an intense mass-transfer phase on a thermal timescale. \cite{1999MNRAS.309..253K,2000ApJ...529..946P,2000MNRAS.317..438K} showed that Cygnus X-2 observations can be explained by case B mass transfer from a donor of mass $3.5$~M$_{\odot}$; that is, the progenitor for the source was an IMXB. 

Contrary to the general understanding of the stability of mass-transfer in X-ray binaries, there is similar evidence in the literature that IMXBs can undergo stable mass transfer with a NS and avoid CE. \cite{2000ApJ...530L..93T} carried out numerical calculations of IMXBs with $2.0$--$6.0$~M$_\odot$ donor and $1.3$~M$_\odot$ accretor masses using an updated version of the Eggleton code \citep{1971MNRAS.151..351E,1972MNRAS.156..361E,1998MNRAS.298..525P}. The authors studied the initial parameter space for producing binary millisecond pulsars with a heavy carbon--oxygen (CO) white dwarf companion, and demonstrated for the first time the full stability of IMXBs. \cite{2009MNRAS.393L..41K} stated that NSs would have lower accretion rates for the same mass-transfer rates because NSs would have stronger beaming as their Eddington limit would be lower than that of BHs. Calculations similar to those of \cite{2000ApJ...530L..93T} were carried out by \cite{2012ApJ...756...85S} using the Eggleton code. These latter authors studied the initial parameter space for binary pulsars while considering orbital angular momentum losses from gravitational wave radiation, magnetic braking, and mass lost from the system. \cite{2017ApJ...846..170T} compared the orbital evolution between IMXBs and HMXBs in studying the connection between ULXs and double NS systems and concluded that the orbital period evolution of IMXBs makes them more likely to be NS ULXs than HMXBs. 

\cite{2015ApJ...802L...5F}   studied the origin of the NS ULX M82~X-2 specifically, by combining parametric population synthesis calculations (using BSE; \citealt{2002MNRAS.329..897H}) with detailed binary evolution calculations (using MESA; \citealt{2011ApJS..192....3P,2013ApJS..208....4P}). Assuming highly non-conservative mass transfer and that a significant fraction of the mass lost from the binary carries the specific orbital angular momentum of the donor star, these latter authors found that the most probable parameters to form a NS ULX are donors with initial masses in the range of $3.0$--$8.0$~M$_{\odot}$ , and initial orbital periods of  $1.0$--$3.0$~days. \cite{2015ApJ...802..131S} suggested that NS ULXs, in general, might represent a higher contribution to the general ULX population than BH ULXs, and a significant portion of those would be IMXBs. Similarly, \cite{2015ApJ...810...20W,2017ApJ...846...17W}, using parametric population synthesis calculations, showed that ULXs are more likely to be NS ULXs than BH ULXs, especially in solar metallicity environments. These latter authors found a typical NS ULX to have a $\sim 1.0$
~M$_{\odot}$ red giant. However, they found that extreme NS ULXs ($L_\mathrm{X}\gtrsim10^{42}$~erg~s$^{-1}$) typically have evolved, low-mass, stripped helium-star donors ($\sim 2.0$~M$_{\odot}$). \cite{2019ApJ...875...53W} found that most BH ULXs emit X-rays isotropically, while NS ULXs are generally beamed. Therefore, even though NS ULXs might be intrinsically more numerous than BH ULXs, these latter authors predict that BH accretors dominate the observed ULX population.

Another discussed property of pulsating ULXs is the very high NS spin-up rate (i.e. the rate of change of spin period) in some observed pulsars. NGC~5907~ULX1 shows a change from a spin period of $1.43$~s to $1.13$~s in about 10 years \citep{2017Sci...355..817I} with an inferred spin-up rate of $-8\pm0.1\times10^{-10}$s~s$^{-1}$. The spin period of NGC~300~ULX1 went from $31.71$~s to $31.54$~s in $4$~days \citep{2018MNRAS.476L..45C} with a spin-up rate $\sim-5.56\times10^{-7}$s~s$^{-1}$, which is the highest rate observed so far for a NS ULX. However, the rate of change of spin period given by  
\begin{equation}
    \dot{P} = \dot{\bigg(\frac{1}{{\nu}}\bigg)} = - \frac{\dot{\nu}}{\nu^2},
\end{equation}
where $\nu$ is the NS spin frequency, does not directly reflect the mass accretion rate. Rather, the rate of change of $\nu$ is roughly proportional to the mass accretion rate \citep[for instance see Eq.~(2) in][]{2017MNRAS.468L..59K}. Nevertheless, we  continue to mention $\dot{P}$ values when talking about observations (Table \ref{tab:obs_data}) as they are the ones most often reported in the literature.

High rates of frequency increase could suggest efficient spin up from very high accretion rates onto the NS. This is because high mass accretion provides the NS with enough torque to spin it up to such high rates \citep{2001ASPC..229..423R, 2006csxs.book..623T}. However, there is a caveat in calculating accretion rates from spin-up frequency as they might not represent the secular average accretion rate over an evolutionary timescale of the donor star. In some cases, the extremely high spin-up rate would grossly overestimate the amount of matter accreted by the NS. For example, M82~X-2 showed a high spin-up rate of $-2\times10^{-10}$s~s$^{-1}$ when X-ray pulses were first discovered \citep{2014Natur.514..202B}. However, later, \cite{2019arXiv190506423B} observed an average spin down of $-5\times 10^{-11}$~Hz~s$^{-1}$ over a period of two years. These latter authors suggested that the source might be close to spin equilibrium and is alternating between phases of spin up and spin down. \cite{2017MNRAS.468L..59K} estimated the spin-up timescale for three pulsating ULXs and concluded that we observe them close to equilibrium. \cite{2019A&A...626A..18C} performed semi-analytical calculations for accretion onto a magnetised NS using the NS being close to spin equilibrium as a boundary condition. The reason for this is that even though $\dot{\nu}$ gives an estimate of the instantaneous accretion rate, it should not be compared directly to long-term average estimates that binary evolution models are giving, as many of these systems might be close to their spin equilibrium.

In our work, we do not investigate X-ray pulses and the super-Eddington emission mechanism. Instead, we focus on ULX formation and long-term evolution. Therefore, in the entirety of this study, we refer to LMXBs/IMXBs with NS accretors that drive super-Eddington mass-transfer rates as NS ULXs. Pulsating ULXs are a subset of NS ULXs as pulsations are not a necessary outcome of super-Eddington mass transfer but rather they are a product of the presence of a relatively strongly magnetised NS. In this paper, we investigate how these NS ULXs could be formed and try to explain the physical properties involved using numerical computations. In doing so, we study how the stability of binaries is affected by spin-orbit coupling, and by a higher accretor mass ($2.0$~M$_\odot$ instead of a typical NS mass of $\sim 1.3$~M$_\odot$). We also assume that there is no precession of the accretion disc or absorption of X-ray flux by optically thick material around the source that might cause the luminosity to vary. In the following section (Section~\ref{sec:observations}), we summarise the properties of the currently observed pulsating ULX sample. In Section~\ref{sec:numericaltools}, we discuss the numerical methods and physics employed for the simulations, while in Section~\ref{sec:results} we present the results from our simulations, highlighting the allowed initial parameter space for NS ULX formation and the properties of the formed population. Section~\ref{sec:discussion} discusses the observed NS ULXs in the context of our results and how the angular momentum exchange between spin and orbit affects the result through tides. Finally, we end with concluding remarks in Section~\ref{sec:conclusion}.

\section{Currently observed pulsating ULX sample}\label{sec:observations}

In this section, we discuss some of the observed and predicted parameters for the NS ULXs present so far in the literature. In most cases, the observables are not well constrained, and therefore large uncertainties are involved and assumptions about the physical properties have been made. In the entirety of this paper we refer to the mass of the donor as $M_{\mathrm{donor}}$, that of the accretor as $M_{\mathrm{acc}}$, the orbital separation as $a$, and the orbital period as $P_{\mathrm{orb}}$. For our work, the most important pulsating ULX is M82~X-2 as it has the most well-constrained observational parameters, even though we comment on other pulsating ULXs observed as well. We summarise all the relevant properties of the currently known sample of pulsating ULXs in Table~\ref{tab:obs_data}.

\begin{table*}[htp]
\begin{center}
\begin{threeparttable}
\begin{tabular}{lllllll}
\hline
NS ULX & $L_\mathrm{X}$ (erg~s$^{-1}$) & $M_{\mathrm{donor}}$ (M$_{\odot}$) & $P_\mathrm{orb}$ (days) & $P_\mathrm{spin}$ (s) & $\dot{P}_\mathrm{spin}$ (s s$^{-1}$) \\ \hline
M82~X-2\tnote{1} & $1.8\times 10^{40}$ & $\gtrsim 5.2$ & $2.52$ & $1.37$ & $-2.0\times 10^{-10}$ \\
NGC~7793~P13\tnote{2} & $5.0\times 10^{39}$ & $18.0$--$23.0$\tnote{3} & $64.0$\tnote{4} & $0.417$ & $-3.5\times 10^{-11}$ \\
NGC~5907~ULX1\tnote{5} & $\sim 10^{41}$ & $2.0$--$6.0$ & $5.3^{+2.0}_{-0.9}$ &  $1.137$ & $-8.1\times 10^{-10}$ \\
NGC~300~ULX1\tnote{6} & $4.7\times 10^{39}$ & $\gtrsim 8.0$--$10.0$\tnote{7} & $\gtrsim 1.0$~yr\tnote{8} &$31.6$ & $-5.56\times 10^{-7}$ \\
M51~ULX-8\tnote{9} & $2.0\times 10^{39}$ & - & - & - & - \\
NGC~1313~X-2\tnote{10} & $1.5\times 10^{40}$ & $\lesssim 12.0$\tnote{11} & - & $1.5$ & $-1.2\times 10^{-10}$ \\
Swift~J0243.6+6124\tnote{12} & $\sim 10^{39}$ & - & $28.3$\tnote{13} & $9.86$\tnote{14} & $-2.2\times 10^{-8}$ \\
M51~ULX-7\tnote{15} & $10^{39}$--$10^{40}$ & $\gtrsim 8.0$ & 2.0 & $2.8$ & $-10^{-9}$ \\
\hline
\end{tabular}
\begin{tablenotes}
\item[1]\cite{2014Natur.514..202B},\item[2]\cite{2017MNRAS.466L..48I, 2016ApJ...831L..14F},\item[3]\cite{2011AN....332..367M},\item[4]\cite{2014Natur.514..198M},\item[5]\cite{2017Sci...355..817I},\item[6]\cite{2018MNRAS.476L..45C},\item[7]\cite{2019arXiv190902171H},\item[8]\cite{2018A&A...620L..12V,2019ApJ...879..130R},\item[9]\cite{2018NatAs...2..312B},\item[10]\cite{2019MNRAS.tmpL.104S},\item[11]\cite{2008A&A...486..151G},\item[12]\cite{2019ApJ...879...61Z}, \item[13]\cite{2018A&A...613A..19D,2017ATel10907....1G},\item[14]\cite{2017ATel10809....1K,2017ATel10812....1J},\item[15]\cite{2019arXiv190604791R}.
\end{tablenotes}
\end{threeparttable}
\end{center}
\caption{Observed and inferred parameters of NS ULXs from the literature. $L_\mathrm{X}$ is the X-ray luminosity, $M_{\mathrm{donor}}$ the donor mass, $P_{\mathrm{orb}}$ the orbital period, $P_{\mathrm{spin}}$ the NS spin period, and $\dot{P}_\mathrm{spin}$ is the time derivative of the NS spin period.}
\label{tab:obs_data}
\end{table*}

\subsection{M82~X-2}\label{m82x2}
This source is one of the most studied NS ULXs. It is observed in the core of M82 galaxy and was discovered to show X-ray pulsations by \cite{2014Natur.514..202B}. It has a peak X-ray luminosity of $1.8\times10^{40}$erg~s$^{-1}$. Using the observed orbital period of $2.52$~days and inclination of $<60\degree$, the binary mass function for M82~X-2 was calculated as $2.1$~M$_{\odot}$. Assuming that the NS mass is $1.4$~M$_{\odot}$, the donor mass was estimated to be $\gtrsim5.2$~M$_\odot$.

\subsection{NGC~7793~P13}\label{7793}
This ULX source appears to be a HMXB containing an accreting NS in galaxy NGC 7793, with a luminosity of about $5.0\times 10^{39}$erg~s$^{-1}$. Earlier, \cite{2011AN....332..367M}  identified the donor in the then-not-discovered ULX system as a late B-type super-giant star with a mass of $18.0$~M$_{\odot}<M_{\mathrm{donor}}<23.0$~M$_{\odot}$.  \cite{2014Natur.514..198M} estimated the orbital period to be 64 days using optical modulation in observations. The NS was found to be present after pulsations were detected from the source \citep{2017MNRAS.466L..48I, 2016ApJ...831L..14F}. 

\subsection{NGC~5907~ULX1}
Discovered to have X-ray pulsations by \cite{2017Sci...355..817I}, this ULX is observed in the galaxy NGC 5907. Its X-ray luminosity, at around $10^{41}$erg~s$^{-1}$, makes it the most luminous ULX discovered to date. This source does not have as well-constrained parameters as M82 X–2. However, some estimates could be made. The authors derived constraints on the orbital period and arrive at a value of $5.3^{+2.0}_{-0.9}$~days using a projected semi-major axis of $a\,\sin i =2.5^{+4.3}_{-0.8}$~lt-s (light-seconds). They also compared  HMXBs, LMXBs, and IMXBs as possible explanations and found them all to be consistent with the upper limits derived from HST images \citep{2013MNRAS.434.1702S}.
% They suggested that it probably is either a HMXB or an IMXB.--deleted

\subsection{NGC~300~ULX1}
This X-ray source, located in the galaxy NGC 300,  was recently discovered to show pulsations \citep{2018MNRAS.476L..45C}. The authors assumed an orbital period in the range $1$--$3$~days and used a projected semi-major axis of $\lesssim4.6\times 10^{-5}$~lt-s and arrived at a binary mass function of $0.8\times 10^{-3}$~M$_{\odot}$. The X-ray luminosity for ULX1 was found to be $4.7\times 10^{39}$erg~s$^{-1}$.
\cite{2019arXiv190902171H} identified the donor as a red super-giant star ($\gtrsim 8.0$~M$_{\odot}$). They reported that the observed effective temperature and luminosity of the donor is consistent with evolutionary tracks of singles stars of initial masses of $8.0$--$10.0$~M$_{\odot}$. \cite{2019MNRAS.488.5225V} observed that the flux of the source decreased by a factor of $\sim 50$ over a few months, while its spin-up remained constant at $4.0\times10
^{-10}$~Hz~s$^{-1}$. This could result from absorption from optically thick material nearby (perhaps from outflows) or from a precessing accretion disc.

\subsection{M51~ULX-8}
\cite{2018NatAs...2..312B} identified M51~ULX-8 as a NS ULX on the basis of the detection of cyclotron resonance scattering features instead of direct observations of X-ray pulses. The cyclotron features observed can be translated into magnetic field strengths, which in this case correspond to that of a highly magnetized pulsar ($\sim 10^{15}$~G). 

\subsection{NGC~1313~X-2}\label{1313}
\cite{2019MNRAS.tmpL.104S} discovered X-ray pulsations in the ULX source NGC~1313 X-2 with a pulse period of $\sim 1.5$~s. \cite{2008A&A...486..151G} estimated the mass of the donor star to be $\lesssim 12.0$~M$_\odot$ assuming it to be part of a metal-poor star cluster with an age of $2.0$~Myr. \cite{2008AIPC.1010..303P} estimated the ULX lifetime using observations of the large bubble nebulae around the source, which is $\sim 1.0$~Myr. Based on studies showing that very few NS ULXs would last as long as $1.0$~Myr \citep[e.g.][]{2017ApJ...846...17W}, \cite{2019MNRAS.tmpL.104S} suggested that NGC~1313~X-2 is in its last stages of mass transfer.

\subsection{Swift~J0243.6+6124}
Discovered by \cite{2017ATel10809....1K} as a transient source, Swift~J0243.6+6124 was known to have an accreting NS with a spin period of $9.86$~s. \cite{2018A&A...613A..19D} estimated an orbital period of $28.3$~days and semi-major axis of $a$ sin$ i =140^{+3}_{-3}$~lt-s. \cite{2019ApJ...879...61Z} studied {\it{Insight}}-HXMT data for this source and estimated its magnetic field as $\sim 10^{13}$~G and an X-ray luminosity of $>10^{39}$erg~s$^{-1}$, confirming the source to be the first Galactic ULX with an NS accretor. Furthermore, these latter authors calculated a high spin-up rate of $-2.2\times 10^{-8}$~s~s$^{-1}$. \cite{2019ApJ...873...19T} reported on the spectral behaviour of this source and suggested that if it were located in an external galaxy it would have a similar appearance to the other pulsating ULXs observed.

\subsection{M51~ULX-7}
\cite{2019arXiv190604791R} discovered $2.8$~s X-ray pulses in observations of ULX-7 in galaxy M51, therefore finding another pulsating ULX following ULX-8 in the same galaxy. The secular spin-up rate was measured as $-10^{-9}$~s~s$^{-1}$ and a variable X-ray luminosity in the range of $10^{39}$--$10^{40}$~erg~s$^{-1}$. The authors used the projected semi-major axis of $28.0$~lt-s and an assumed accretor mass of $1.4$~M$_{\odot}$ to infer a donor of mass $>8.0$--$13.0$~M$_{\odot}$. \cite{2020MNRAS.491.4949V} studied the X-ray light curves of the source and estimated a magnetic field of the NS (rotating near spin equilibrium) of $2.0$--$7.0\times 10
^{13}$~G. They also assumed that the NS is freely precessing and estimated the magnetic field to be $3.0$--$4.0\times 10
^{13}$~G, agreeing with their previous estimate.

%%%%%%%%%%%%%%%%%%%%%%%%%%%%%%%%%%%%%%%%%%%%%%%%%%%%%%%%%%%%%%%%%%%%%%%%%%%%%%%%%%%
\section{Numerical tools and calculations}\label{sec:numericaltools}

In this section we discuss the numerical code used for the binary evolution calculations along with the adopted model parameters and the code modifications that we introduced. 

\subsection{Numerical stellar evolution code and progenitor binary}
\label{code}

To simulate the evolution of the binaries we use MESA (version 10108; MESASDK version 20180127) which is a stellar structure and binary evolution code developed by \cite{2011ApJS..192....3P,2013ApJS..208....4P,2015ApJS..220...15P,2018ApJS..234...34P,2019ApJS..243...10P}.

All LMXBs/IMXBs are calculated as systems with an initially zero-age main-sequence (ZAMS) donor and a point-mass NS accretor. It is assumed that the NS was formed during a previous evolutionary stage which we do not study, and the binary survived a possible NS natal kick from the supernova explosion that formed the NS. We compute a grid of models spanning $0.92$--$8.0$~M$_{\odot}$ in initial donor mass, and $0.5$--$100$~days in initial orbital period. For reference, initial parameters refer to the orbital parameters at the onset of RLO.
The calculation of the mass-transfer rate during RLO is done implicitly using the scheme proposed by \cite{1990A&A...236..385K}.

We use $1.3$~M$_{\odot}$ and $2.0$~M$_{\odot}$ for the mass of the accretor;  $1.3$~M$_{\odot}$ is close to the post-supernova Chandrasekhar mass limit for the formation of a NS and $2.0$~M$_{\odot}$ is on the high-mass end of the NS mass distribution \citep{2010arXiv1012.3208L,2013Sci...340..448A, 2018ApJ...852L..25R}. The NS companion to the pulsar J0453+1559 has a mass of $1.174^{0.004}_{0.004}$~M$_{\odot}$ \citep{2015ApJ...812..143M}\footnote{See \cite{2019ApJ...886L..20T} for an alternative possibility} and therefore NS masses below $1.3$~M$_{\odot}$ have indeed been observed. However, we take $1.3$~M$_{\odot}$ as the standard NS mass.

For the radiative efficiency of the accretion onto the NS with initial mass $M^i_{\rm acc}$ (i.e. the release of gravitational energy of the infalling material in the form of radiation; in units of rest-mass energy) we use,

\begin{equation}\label{eta}
    \eta = \frac{G M^i_{\rm acc}}{c^2 R_{\rm acc}},
\end{equation}
where, $c$ is the speed of light and $R_{\rm acc}$ is the NS radius which we take as 11.0~km.
For the Eddington limit of an accretor with initial mass $M^i_{\rm acc}$,
\begin{equation}\label{Ledd}
    L_{\rm Edd} = \frac{4\pi G M^i_{\rm acc} c }{\kappa},
\end{equation}
where, $\kappa$ is the opacity. Using Eq.~(\ref{eta}) and simplifying Eq.~(\ref{Ledd}) (for accretion of pure ionised hydrogen), we get
\begin{equation}\label{Eq:Edd}
    \dot{M}_\mathrm{Edd}=1.5\times 10^{-8}~\bigg(\frac{M^i_{\mathrm{acc}}}{1.3~ \mathrm{M}_{\odot}}\bigg) ~\mathrm{M}_{\odot}~\mathrm{yr}^{-1}.
\end{equation}

We fix the Eddington limit and the radiative efficiency to initial values as they would not change significantly from the amount of mass that the NSs accrete in our grids. If the mass-transfer rate goes beyond this value, the radiation pressure will prevent any excess material from being accreted. Mass from the donor is transferred conservatively to the accretor and a fraction of this transferred mass is lost from the vicinity of the accretor as an isotropic fast wind or jet with the specific angular momentum of the accretor \citep[see][for a detailed explanation]{2006csxs.book..623T}. The efficiency of accretion ($\epsilon$) by the NS is $\epsilon = 1 - (\alpha + \beta + \delta)$, where $\alpha$ is the fractional mass lost directly from the donor, $\beta$ is the fractional mass lost from the vicinity of the accretor, and $\delta$ is the fractional mass lost from a circumbinary toroid. We take the values, $\alpha =\delta= 0$, and $\beta = {\it max}~\{0.7, 1-{\dot{M}_{\mathrm{Edd}}}/{\dot{M}_{\mathrm{donor}}}\}$. 

We consider orbital angular momentum (J$_{\mathrm{orb}}$) losses via gravitational wave radiation, spin-orbit coupling due to tidal effects (Section~\ref{sync_time}), and mass lost from the system. We also include effects due to magnetic braking following the prescription by \cite{1983ApJ...275..713R} for donor masses that develop an outer convective envelope at any point. We take the eccentricity to be negligible, as tidal forces would circularise the orbit of a semi-detached binary with a giant star on a relatively short timescale of $10^4$~years \citep{1995A&A...296..709V}. This is orders of magnitude shorter than the main sequence lifetime of intermediate-mass stars which are of the order of $10^8$--$10^9$~years (main sequence lifetime of low-mass stars would be even longer). Furthermore, tidal forces aim to synchronise the stars with the orbit. We assume the orbit is synchronised by the time RLO begins since the tidal forces would cause the stars to synchronise with the orbit on a relatively short timescale \citep{2000ScChA..43..331H}.

We assume solar metallicity, that is Z$_{\odot}$=0.02, and that any layer in the donor interior is stable against convection if the Ledoux criteria for convection is fulfilled \citep{1947ApJ...105..305L}. At the edges of convective zones, we account for overshooting because the convective material slightly enters non-convective zones due to inertia. To describe overshooting we follow the exponential overshooting efficiencies used in the MIST models for low- and intermediate-mass stars, which are $f_{\mathrm{ov, core}} = 0.0160$ in the core calibrated from properties of the open cluster M67 and $f_{\mathrm{ov, en}} = 0.0174$ in the envelope calibrated from solar properties \citep{2016ApJS..222....8D, 2016ApJ...823..102C}. For stellar winds, we use the cool red giant branch wind scheme described by \cite{1975psae.book..229R} with a scaling factor of 0.1. In cases where
we get a stripped helium (He) star, we use the prescription for Wolf-Rayet stars by \cite{2000A&A...360..227N}, included in MESA under the Dutch hot wind scheme. 

The mass-transfer calculations are carried out until one of the following conditions are met: \textit{(i)} the donor forms a white dwarf (WD), \textit{(ii)} the age of the donor star  exceeds the Hubble time, \textit{(iii)} the radius of the donor star extends so far beyond its Roche lobe that L$_{2}$ overflow is initiated and the system becomes dynamically unstable (see Section~\ref{rl2}), \textit{or (iv)} the number of computational steps exceeds a limit of 300,000 (this value was chosen based on previous grid runs). We include condition \textit{(iv)} for those systems where MESA runs into converging problems and cannot find a solution. This happens for only two types of binaries in our numerical calculations. In the first case, the mass transfer cannot properly remove the last bit of the envelope from the donor. In the second case, MESA is not able to solve for the donor radius which extends quite far beyond the Roche lobe but not enough to trigger the condition of L$_2$ overflow. Both these cases occur at the end of the mass-transfer phase, and therefore we accept that the binary was stable until the end of its evolution.

\subsection{Super-Eddington accretion onto the NS}\label{acc}

For non-conservative super-Eddington mass transfer the amount of mass that is accreted is less than that transferred per unit time ($0.3\times $ mass transferred, as per our assumptions) until the Eddington limit prevents a further increase in accretion rate. Cygnus X-2 is an example of a system observed to have survived super-Eddington mass transfer while presumably having accreted comparatively less. This source suggested that high mass transfer onto a NS (or BH) could avoid a CE phase with the NS ejecting most of the mass that was transferred at super-Eddington rates \citep{1999MNRAS.309..253K, 1999ApJ...519L.169K, 2000ApJ...529..946P, 2001ApJ...552L.109K}.

\cite{1973A&A....24..337S} studied the observational characteristics of accretion discs in sub-Eddington and super-Eddington regimes of mass transfer in the case of black-hole binaries. In this picture, as the mass transfer approaches the  Eddington limit, the structure of the disc changes from a slim disc to a disc with an inner geometrically thick component and an outer thin component. As matter that is transferred to the accretion disc at super-Eddington rates moves radially inwards (transporting angular momentum outwards), strong outflows begin at a certain radius which remove a fraction of the matter, thereby also taking away excess angular momentum \cite[see][for an application of the super-Eddington disc model]{2013AstBu..68..139V}. The spherisation radius is defined as the radius at which the accretion luminosity first reaches the Eddington limit and strong outflows begin, and one can approximate it as 
\begin{equation}
     R_\mathrm{sph}= \dot{M}_{\mathrm{donor}}\frac{G M_\mathrm{acc}}{L_\mathrm{Edd}}.
\end{equation}
Outside $R_\mathrm{sph}$ the disc emits X-rays with luminosity $L_\mathrm{Edd}$ which depends on the accretion rate following the equation,
\begin{equation}
    L_\mathrm{Edd}= \eta \dot{M}_{\mathrm{Edd}}c^2.
\end{equation}

Inside $R_\mathrm{sph}$, the outflowing matter has a velocity which depends on the difference between inward gravity and outward radiation pressure. The velocity of the outflow increases inward which in turn decreases the mass-accretion rate at each radius. This keeps the disc locally Eddington limited. The mass-accretion rate within $R_{\mathrm{sph}}$ at each point in the disc can be described by the following equation:
\begin{equation}
    \dot{M}^{\mathrm{local}}_{\mathrm{acc}}(R) = \dot{M}_{\mathrm{donor}} \frac{R}{R_\mathrm{sph}}.
\end{equation}
Since the radiation pressure is balanced by the gravitational pressure at each point, the accretion disc inside the spherisation radius has a thickness of the order of the distance from the accretor and emits X-rays with luminosity $L_\mathrm{Edd}\times\ln{\dot{m}}$ \citep[where 
$\dot{m} \equiv \dot{M}_\mathrm{donor}/\dot{M}_\mathrm{Edd}$; ][]{1973A&A....24..337S}. For $\dot{m}>1$, the total radiated luminosity from the accretor can exceed the Eddington limit by
\begin{equation}\label{edd_eq}
    L_{\mathrm{acc}} = L_{\mathrm{Edd}}(1 + \ln{\dot{m}}).
\end{equation}

For highly super-Eddington mass-transfer rates another effect could come into play because of the geometrically thick accretion disc.
A narrow funnel forms along the rotation axis of the accretor from where radiation can escape as a collimated jet. The observed isotropic-equivalent accretion luminosity as described by \cite{2001ApJ...552L.109K} and \cite{2009MNRAS.393L..41K} is then as follows:

\begin{equation}\label{new_edd}
    L^{\mathrm{iso}}_{\mathrm{acc}} = \frac{L_{\mathrm{Edd}}}{b}(1 + \ln{\dot{m}}),
\end{equation}
where $b$ is the beaming factor describing the amount of collimation to the outgoing radiation. The approximated value of $b$ is
\begin{equation}\label{eq:beaming}
        b= 
\begin{dcases}
    \frac{73}{\dot{m}^2},& \text{if } \dot{m}> 8.5,\\
    1,              & \text{otherwise}.
\end{dcases}
\end{equation}
Because the beaming factor is an approximation, we apply an upper limit ($\sim 10^{42}$~erg~s$^{-1}$) in calculating the isotropic-equivalent accretion luminosities so that we do not get unphysically high values.

\subsection{Spin-orbit coupling and synchronisation timescales}
\label{sync_time}

As mentioned before, tidal forces synchronise the stars with the orbit on a timescale which is relatively short. Therefore, whenever there is a change in either the spin angular momentum or the orbital angular momentum, tides will work to synchronise the system again. This action of tides on the orbit may affect the stability and evolution of the binary \citep{2001nsbh.conf..337T}.

When mass is lost from the donor star during mass transfer it also removes spin angular momentum from the star which is supplied to the orbit. If the spin angular momentum removed from the donor and returned to the orbit is non-negligible, then it has a widening effect on the orbit. This is followed by mass loss from the accretor's vicinity, with the mass lost carrying away the  specific angular momentum of the accretor. This competing effect tends to shrink the orbit. Mass that is leaving the system from the accretor's vicinity carries with it relatively high specific orbital angular momentum (when the accretor is the less massive binary component), having a shrinking effect on the orbit. The donor then becomes sub-synchronous with the orbit and angular momentum has to be transferred from the orbit to the star in order to spin it up, causing the orbit to shrink even further. The interplay between the orbital shrinking and widening effects can reveal how spin-orbit coupling affects the stability of mass transfer. In the presence of strong winds, there is additional loss of angular momentum from the donor. However, for stars with masses of $0.92$--$8.0$~M$_{\odot}$, stellar winds are too weak to cause significant orbital change. 

The tidal synchronisation timescale is defined as follows \citep{1977A&A....57..383Z,1981A&A....99..126H}:
\begin{equation}
    \frac{1}{T_\mathrm{sync}} = 3 \frac{K}{T} \bigg(\frac{q}{r_\mathrm{g}}\bigg)^2 \bigg(\frac{R_\mathrm{donor}}{a}\bigg)^6,
\end{equation}
where $q$ is the mass ratio (we define $q\equiv M_{\mathrm{acc}}/{M_{\mathrm{donor}}}$) and $r_\mathrm{g}$ is the gyration radius of the star. ${K}/{T}$ is the spin-orbit coupling parameter and is described in two ways for a star with an outer radiative envelope as $({K}/{T})_{\mathrm{rad}}$, and a star with an outer convective envelope as $({K}/{T})_{\mathrm{conv}}$. For the former case we use,
\begin{align}
    \bigg(\frac{K}{T}\bigg)_{\mathrm{rad}} &= \bigg(\frac{GM_{\mathrm{donor}}}{R_\mathrm{donor}^3}\bigg)^{1/2}(1+q)^{5/6}E_{2}\bigg(\frac{R_\mathrm{donor}}{a}\bigg)^{5/2},\\
    E_2 &= 10^{-0.42}\bigg(\frac{R_\mathrm{conv}}{R_\mathrm{donor}}\bigg)^{7.5},
\end{align}
where $R_\mathrm{conv}$ is the radius of the convective core. The value for $E_2$ is computed from fitting formulae for H-rich stars derived by \citet{2018A&A...616A..28Q}. For $({K}/{T})_{\mathrm{conv}}$, the definition is taken from \cite{2002MNRAS.329..897H} to be
\begin{equation}
    \bigg(\frac{K}{T}\bigg)_{\mathrm{conv}} = \frac{2}{21}\frac{f_\mathrm{conv}}{\tau_\mathrm{conv}}\frac{M_\mathrm{env}}{M_{\mathrm{donor}}} \mathrm{yr}^{-1},
\end{equation}
where $f_\mathrm{conv} = {\it min}~\{1.0, (P_{\rm tidal}/(2\tau_{\rm conv}))^{2}\}$ is a numerical factor and $P_{\rm tidal}$ is the tidal pumping time-scale defined by $\lvert 1/P_{\rm orb} - 1/P_{\rm spin}\rvert$. Here, $\tau_\mathrm{conv}\approx(MR^2/L)^{1/3}$ is the eddy turnover timescale in units of years \citep{1996ApJ...470.1187R} and $M_\mathrm{env}$ is the convective envelope mass. More details about our applied orbital angular momentum evolution, with spin-orbit coupling including tides, are discussed in Section~\ref{ls-coup} and Appendix~\ref{append1}.

\subsection{Angular momentum loss from the second Lagrangian point}{\label{rl2}}

\begin{figure}[!ht]
\centering
\includegraphics[width=\linewidth]{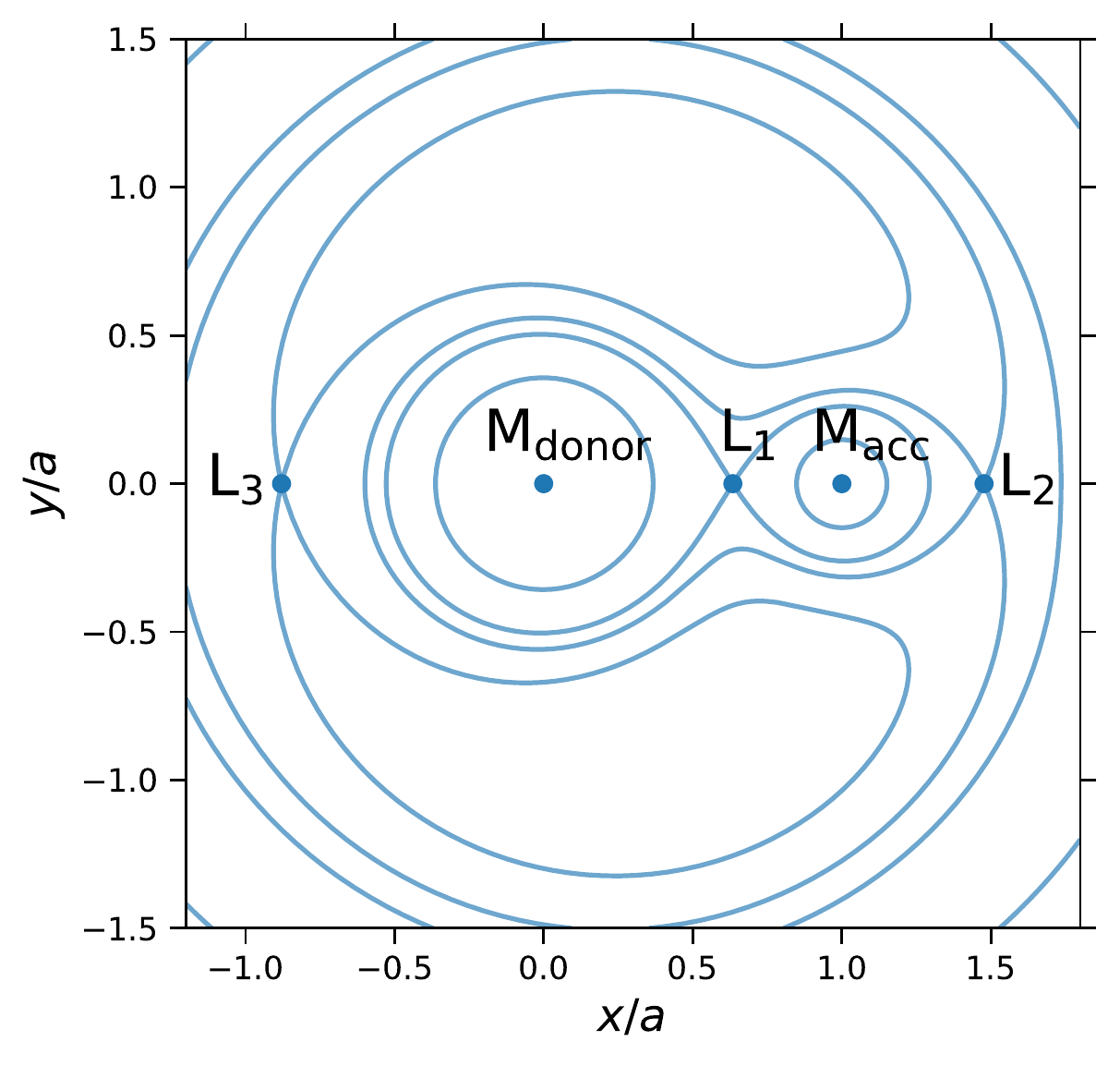}
\caption{Equipotential lines of the Roche potential for a binary consisting of stars with mass ratio $q=0.26$ and binary separation $a$. The equipotential lines passing through Lagrangian points L$_1$, L$_2$ , and L$_3$ are shown.}
\label{contour}
\end{figure} 

\begin{figure}[ht]
\centering
\includegraphics[width=\linewidth]{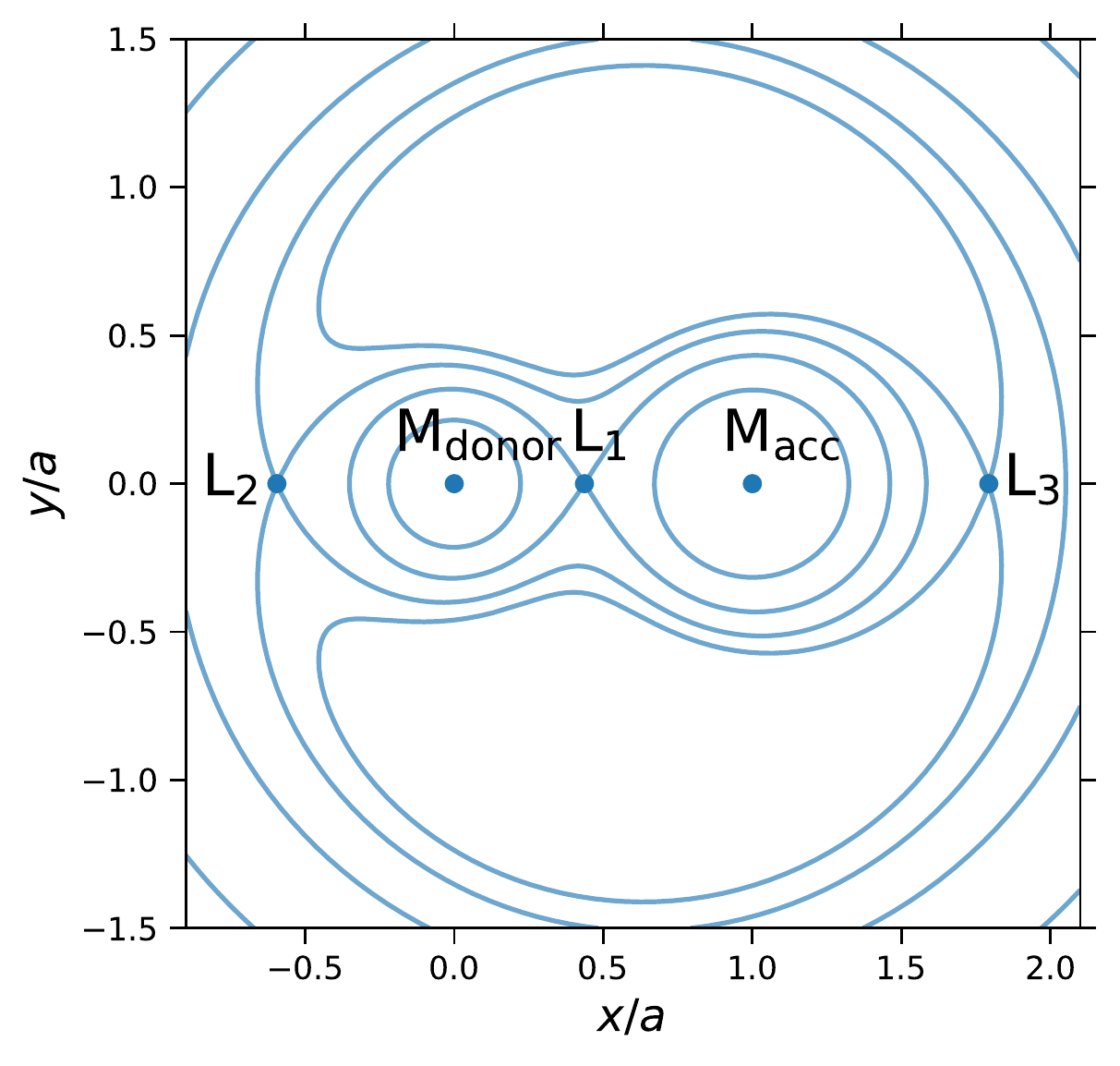}
\caption{Equipotential lines of the Roche potential for a binary consisting of stars with mass ratio $q=1.85$ and binary separation $a$ to illustrate the swapping of the positions of L$_2$ and L$_3$ when $M_{\mathrm{donor}}\leq M_{\mathrm{acc}}$ (compared to Fig.~\ref{contour}).}
\label{contour_f}
\end{figure} 

\begin{figure*}[ht]
\centering
\includegraphics[scale=0.6]{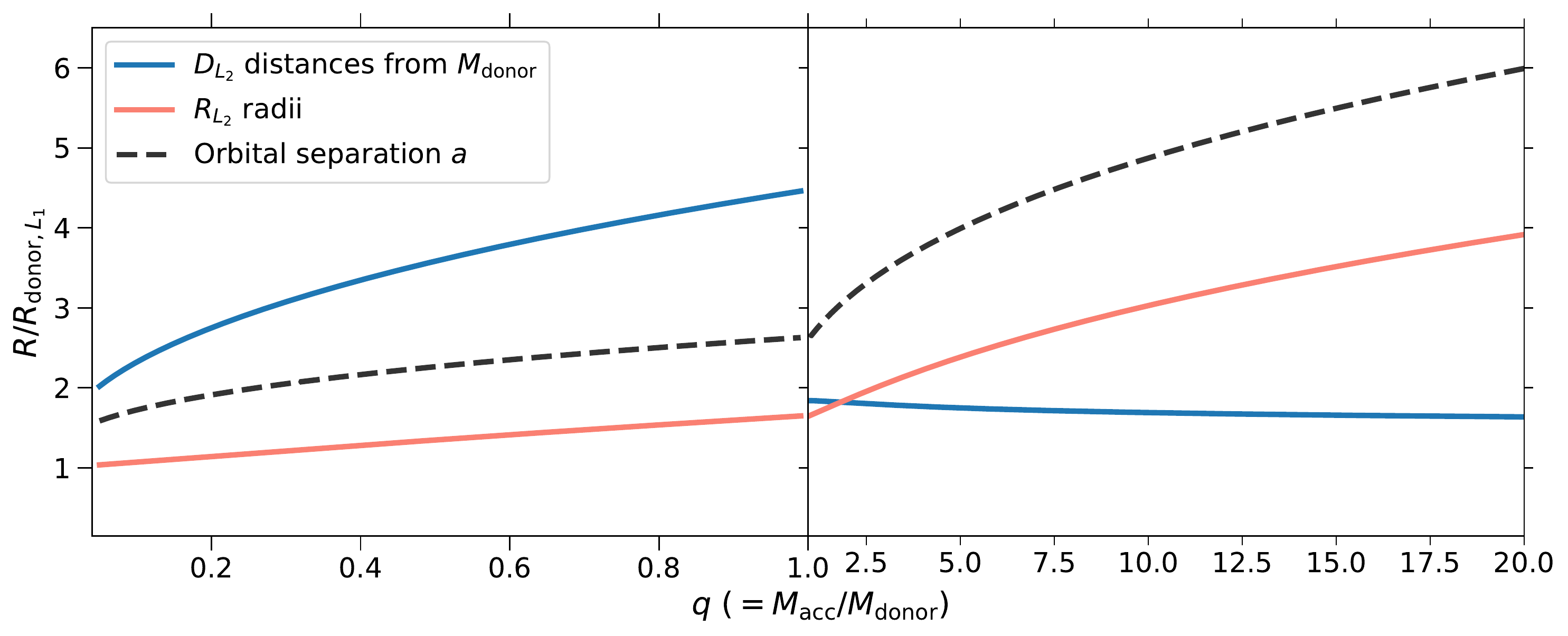}
\caption{Our results form the volume integration of L$_2$ equipotential surfaces with respect to different mass ratios $q$. All distance values are presented in units of $R_{\mathrm{L}_{1}}$. Here, $R_{\mathrm{L}_2}$ is the radius of a sphere with volume equal to that of the L$_2$ equipotential surface (solid orange line). $D_{\mathrm{L}_2}$ is the distance of the L$_2$ point from $M_{\mathrm{donor}}$ (solid blue line). The dashed black line shows the orbital separation in units of $R_{\mathrm{L}_{1}}$. Systems where the donor's radius exceeds any of these limiting radii are considered as undergoing dynamical instability.}
\label{L2_R2}
\end{figure*} 

Lagrangian points are equilibrium points in space where the gravitational and centrifugal forces in the system cancel each other out. L$_1$, L$_2$, and L$_3$ are unstable equilibrium points from where a test particle, upon small displacement, would move further away. Figure~\ref{contour} shows these unstable Lagrangian points for a system with a mass ratio of $q=0.26$.

In most cases of X-ray binaries, analysis has been done for mass transfer via L$_{1}$. L$_1$ lies in between the two stars (hence also known as the inner Lagrangian point) and the equipotential surface passing L$_1$ is known as the Roche lobe. When a star fills its Roche lobe, any material that crosses L$_1$ from one star will fall towards the other star. This transfer of matter either decreases the radius (for radiative envelopes) of the donor or increases it (for convective envelopes). In some cases of extreme binary mass ratio, the RLO might not be enough to provide efficient mass-transfer rate and the donor might extend far beyond its Roche lobe to reach the equipotential surface passing through L$_2$. However, contrary to when mass passes through L$_1$, the mass that crosses L$_2$ takes away a large amount of angular momentum from the binary. Once the outer layers of the donor reach L$_2$ (or the donor obtains a volume equivalent to that of the equipotential lobe passing through L$_2$; see below) it is expected that the binary orbit will shrink rapidly. We consider this the onset of dynamical instability. For illustration, the L$_2$ potential surface is shown in Fig.~\ref{contour}. It is the peanut-shaped surface enclosing both binary mass components and passing through the point L$_2$. 

\cite{1983ApJ...268..368E} calculated the stellar radius needed in order to initiate mass transfer from the inner Lagrangian point of a binary by calculating the radius of a sphere that will have the same volume as the Roche lobe. This radius for the donor star is referred to as the Roche-lobe radius ($R_{\mathrm{donor},\mathrm{L}_1}$). We take a similar approach to quantify overflow from L$_2$. In cases where mass transfer via the L$_1$ point is not sufficient to keep the donor star confined within its Roche lobe we assume that the expanding star needs to fill the entire volume enclosed by the L$_2$ equipotential surface before the onset of dynamical instability. $R_{\mathrm{L}_2}$ is the volume equivalent radius for the equipotential surface passing through the L$_{2}$ point.

Another possibility of mass loss from L$_2$ point occurs when the radius of the donor star reaches the point L$_2$ before the donor volume overfills the L$_2$ potential surface. This case applies only when $q\ge 1$ (i.e. $M_{\mathrm{donor}}\le M_{\mathrm{acc}}$) as the point L$_2$ is much closer to the donor. We refer to the distance between the centre of the donor and L$_2$ as $D_{\mathrm{L}_2}$. This case is illustrated in Fig.~\ref{contour_f} which shows the equipotential surfaces  with mass ratio, $q=1.85$. In our simulations we assume that binaries experience stable mass transfer when the donor star does not cross any of the two limits discussed above at any point during each evolution (i.e. for stable RLO: $R_{\rm donor}< {\it min}~\{R_{\mathrm{L}_2},~D_{\mathrm{L}_2}\}$ for the entire binary evolution).

To calculate the L$_2$ volume via numerical integration, we begin by finding the Lagrange points (L$_1$, L$_2$ and L$_3$) for a particular mass ratio. Along the axis joining the centres of the $M_{\mathrm{donor}}$ and $M_{\mathrm{acc}}$, the entire volume is assumed to be a summation of thin discs. The volume of each disc slice is calculated going from one boundary end to the other using the boundaries of the L$_2$ equipotential surface, and the subsequent disc volumes are added together to cover the entire volume. In these calculations we consider two mass ratio regimes, $q < 1$ and $q\ge 1$.

We use both $R_{\mathrm{L}_2}$ and $D_{\mathrm{L}_2}$ normalised to R$_{\mathrm{donor},\mathrm{L}_1}$ in order to remove the dependence on the binary orbital separation. Once these radii are calculated we find a fit of ${R_{\mathrm{L}_2}}/{R_{\mathrm{donor},\mathrm{L}_1}}$ and ${D_{\mathrm{L}_2}}/{R_{\mathrm{donor},\mathrm{L}_1}}$ on the mass ratio ($q=M_{\rm acc}/M_{\rm donor}$) of the binary. For $q < 1$, we find that the $R_{\mathrm{L}_2}$ and $D_{\mathrm{L}_2}$ follow monotonically increasing trends toward $q = 1$ (shown in Fig.~\ref{L2_R2} in the left panel), which is fitted by the following functions,
\begin{equation}\label{eq1}
    \frac{R_{\mathrm{L}_2} (q < 1)}{R_{\mathrm{donor},\mathrm{L}_1}} = 0.784~q^{1.05} e^{-0.188~q} + 1.004,
\end{equation}
\begin{align}\label{eq2}
    \frac{D_{\mathrm{L}_2}(q < 1)}{R_{\mathrm{donor},\mathrm{L}_1}} = 3.334~q^{0.514} e^{-0.052~q} + 1.308.
\end{align}
Here, $R_{\mathrm{donor},\mathrm{L}_1}$ is also calculated from the volume of the L$_{1}$ equipotential surface using the method described above and the result is consistent with the calculations from \cite{1983ApJ...268..368E}. 

For the second mass ratio regime, we calculate $R_{\mathrm{L}_2}(q\ge 1)$  using

\begin{equation}
    R_{\mathrm{L}_2}(q) = R_{\mathrm{L}_2}\bigg(\frac{1}{q}\bigg) \qquad \forall\; q>0.
\end{equation}
We find $D_{\mathrm{L}_2}(q\ge 1)$ using the distance between L$_2$ and $M_{\mathrm{acc}}$ from the case $q < 1$. As seen in Fig.~\ref{L2_R2} on the right, there is a sudden jump in the values of $D_{\mathrm{L}_2}$: when crossing $q=1,$ it goes from around $4.45$ for $q=0.997$ to around $1.82$ for $q=1.003$. We fitted functions to the calculated $R_{\mathrm{L}_2}$ and $D_{\mathrm{L}_2}$ values as follows,
\begin{equation}\label{eq3}
    \frac{R_{\mathrm{L}_2}(q \ge 1)}{R_{\mathrm{donor},\mathrm{L}_1}} = 0.290~q^{0.829}~e^{-0.016~q} + 1.362,
\end{equation}
\begin{equation}\label{eq4}
    \frac{D_{\mathrm{L}_2}(q \ge 1)}{R_{\mathrm{donor},\mathrm{L}_1}} = -0.040~q^{0.866}~e^{-0.040~q} + 1.883.
\end{equation}

The relative errors between our calculations and the corresponding fits are less than 1\%. 

Volume equivalent L$_{2}$ radii calculations have also been done by \cite{2016A&A...588A..50M} for $q<1$ but with an entirely different approach as these latter authors considered the case of overcontact binaries. They calculated the L$_{2}$ volume assuming that both the stars expand and fill their respective L$_2$ sub-volume. In order to test our numerical volume-integrating scheme against that of \cite{2016A&A...588A..50M}, we split the L$_2$ volume at the L$_1$ point and compared the resulting calculations to their $R_{\mathrm{L}_{2}}$~radii, finding good agreement. However, we should stress again that the approach by \cite{2016A&A...588A..50M} is only applicable to overcontact binaries. Our work was followed by \cite{2020arXiv200600774G} who, in a different context, derived fits to the volume equivalent L$_{2}$ radius using a similar approach.

%%%%%%%%%%%%%%%%%%%%%%%%%%%%%%%%%%%%%%%%%%%%%%%%%%%%%%%%%%%%%%%%%%%%%%%

\section{Results}\label{sec:results}

We explore the evolution of LMXBs/IMXBs with different initial conditions,  taking into account the physics described in the earlier sections. In both our grids (for NS masses of 1.3~M$_{\odot}$ and 2.0~M$_{\odot}$ respectively) the binaries that interacted via mass transfer can  undergo either stable or unstable mass transfer, excluding the systems with P$^i_\mathrm{orb} \lesssim 0.50$~days where the donor star already overflows its Roche lobe at ZAMS which we do not further evolve. The superscript $i$ stands for initial values which corresponds to the orbital parameters at the onset of RLO. We flag systems as `stable' when either the donor has detached from its Roche lobe at the end of mass transfer or the hydrogen in the outer layer has been almost completely removed (remaining hydrogen in the outer layer $<0.005$~M$_\odot$). Donors in stable binaries formed a WD at the end of the mass-transfer sequence resulting in a neutron star--white dwarf (NS--WD) binary. The `unstable' binaries underwent the onset of L$_{2}$ overflow.

\begin{figure}[t]
    \vfill
        \includegraphics[width=\linewidth]{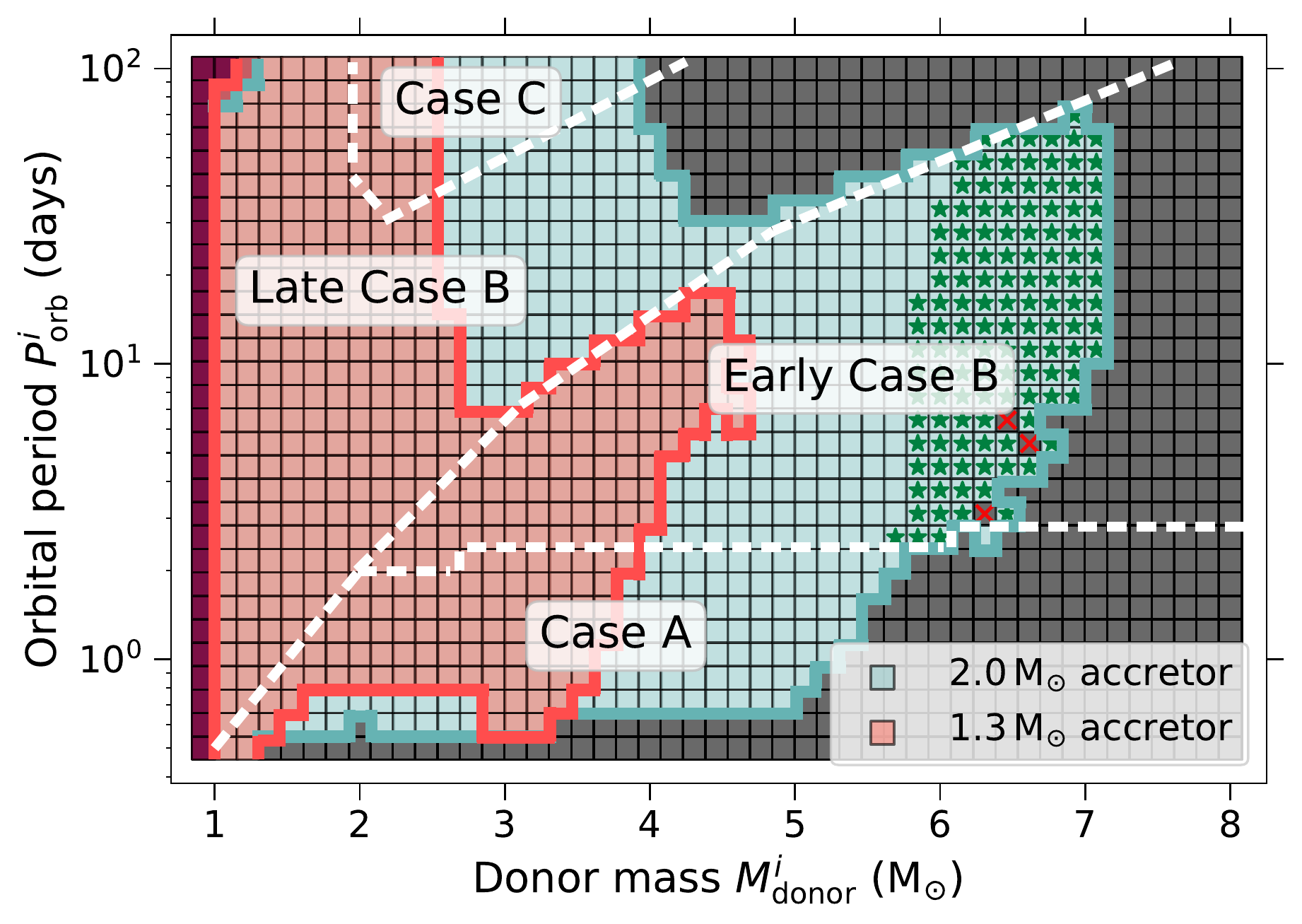}
        \caption{Allowed initial parameter space for LMXBs/IMXBs to undergo stable mass transfer with $1.3$~M$_\odot$ accretors (orange squares) and $2.0$~M$_\odot$ accretors (blue squares). Grey squares correspond to systems that encountered dynamical instability. The dark red squares towards the left edge of the figure correspond to systems that never initiated RLO. The lower dashed white line separates systems undergoing case A mass transfer from those undergoing case B mass transfer (same boundary in both grids). The middle dotted white line separates early case B (where the donor has yet to form a deep convective envelope) from late case B RLO (where donor has formed a deep convective envelope at onset of RLO). The upper dashed white curve encloses systems that undergo case C mass transfer (same boundary in both grids). Green stars correspond to those systems that undergo a second mass-transfer phase from a stripped Helium-giant star, i.e. case BB RLO. The three red crosses in case BB correspond to systems that terminated due to numerical issues.}
        \label{rlo_cases1}
\end{figure}

In Fig.~\ref{rlo_cases1} we present this allowed initial parameter space for LMXBs/IMXBs to undergo stable mass transfer for grids containing $1.3$~M$_\odot$ (orange squares) and $2.0$~M$_\odot$ NS accretors (blue squares) along with the unstable sequences (grey squares). The dark red squares towards the left are the systems that never initiated RLO. The general shape of the stable region resembles the work done by \cite{2000ApJ...530L..93T} where they explored the allowed parameter space to form binary millisecond pulsars while avoiding a CE phase. The different dashed white lines separate the grids based on the type of mass-transfer phase that the system underwent. Case A is when the donor is on the main sequence at the onset of mass loss, that is, it is burning hydrogen in the core. Case B is when the donor has exhausted H in its core and H-shell burning phase (post-main sequence). The threshold between the two cases A and B (lower dashed white line in figure) depends more on the initial orbital period than the donor mass; the limit for case A being in the range $2.0$--$2.8$~days for both grids. The middle dashed white line separates two subsets of case B RLO: early case B and late case B. This threshold depends almost linearly on both the initial orbital period and the initial donor mass. At RLO onset, if the donor has a radiative envelope it is termed early case B and if the donor has developed a convective envelope it is termed late case B RLO. The upper dashed white curve encloses systems that undergo case C RLO, which means the donor has exhausted He in its core at the onset of RLO. 

The stability region increases for higher accretor mass to include higher donor masses and orbital periods. A similar effect of increase in the parameter space with increasing the accretor mass was obtained by \cite{2012ApJ...756...85S} for the formation of recycled pulsars from IMXBs and LMXBs. The overall shape of the parameter space for both the grids depends a lot on the structure of the donor envelope and the response of the donor to mass loss. A radiative envelope would shrink on mass loss and contribute to the stability of the binary while a convective envelope would expand rapidly on mass loss and make the system increasingly unstable. \cite{1999ApJ...519L.169K} showed that mass transfer on a thermal timescale would avoid the CE phase in a binary as long as the envelope is mostly radiative. Case A and early case B have radiative envelopes, with binaries of initial mass ratio greater than $\sim 0.28,$ undergoing stable mass-transfer depending on the initial orbital period. In contrast, late case B and case C have convective envelopes at RLO with the stability region being defined by a fixed critical mass ratio which we find to be at around $\sim 0.5$. \cite{2014A&A...571A..45I} studied the binary evolution in close LMXBs and also found an increase in the depth of the convective envelope with increasing P$^i_\mathrm{orb}$ (Fig.~4 in their paper).

In Fig.~\ref{rlo_cases1}, green stars correspond to systems that go through case~BB RLO, which is when a binary, after having lost its hydrogen envelope in a case~B mass-transfer phase, detaches and evolves as a stripped helium star and initiates a second RLO phase during the helium-shell burning stage. This subset of case B occurs for only a small part of the parameter space (and only for the grid with a $2.0$~M$_{\odot}$ accretor). This is because low-mass helium stars ($<0.8\;M_\odot$) do not expand by any significant amount  \citep[e.g.][]{heu16,2018MNRAS.481.1908K} and thus only donor stars $\ga 5.8\;M_\odot$ leave behind stripped helium stars massive enough to eventually lead to case~BB RLO. The three red crosses in case BB correspond to systems that terminated due to numerical issues.

During the RLO phase, all LMXBs/IMXBs are expected to be bright X-ray binaries, and in many cases the mass-transfer rate from the donor star to the vicinity of the NS can exceed the Eddington limit ($L_{\rm Edd}$) significantly. As described in Section~\ref{acc}, we only allow accretion onto the NS up to the Eddington limit, but we do consider the transition from a thin accretion disc to an inner thick disc model when the mass-transfer rate supplied from the donor star exceeds the Eddington limit \citep{1973A&A....24..337S}, as well as the geometric beaming model proposed by \cite{2001ApJ...552L.109K}. 

\begin{figure}[t]
    \vfill
    \centering
        \includegraphics[width=\linewidth]{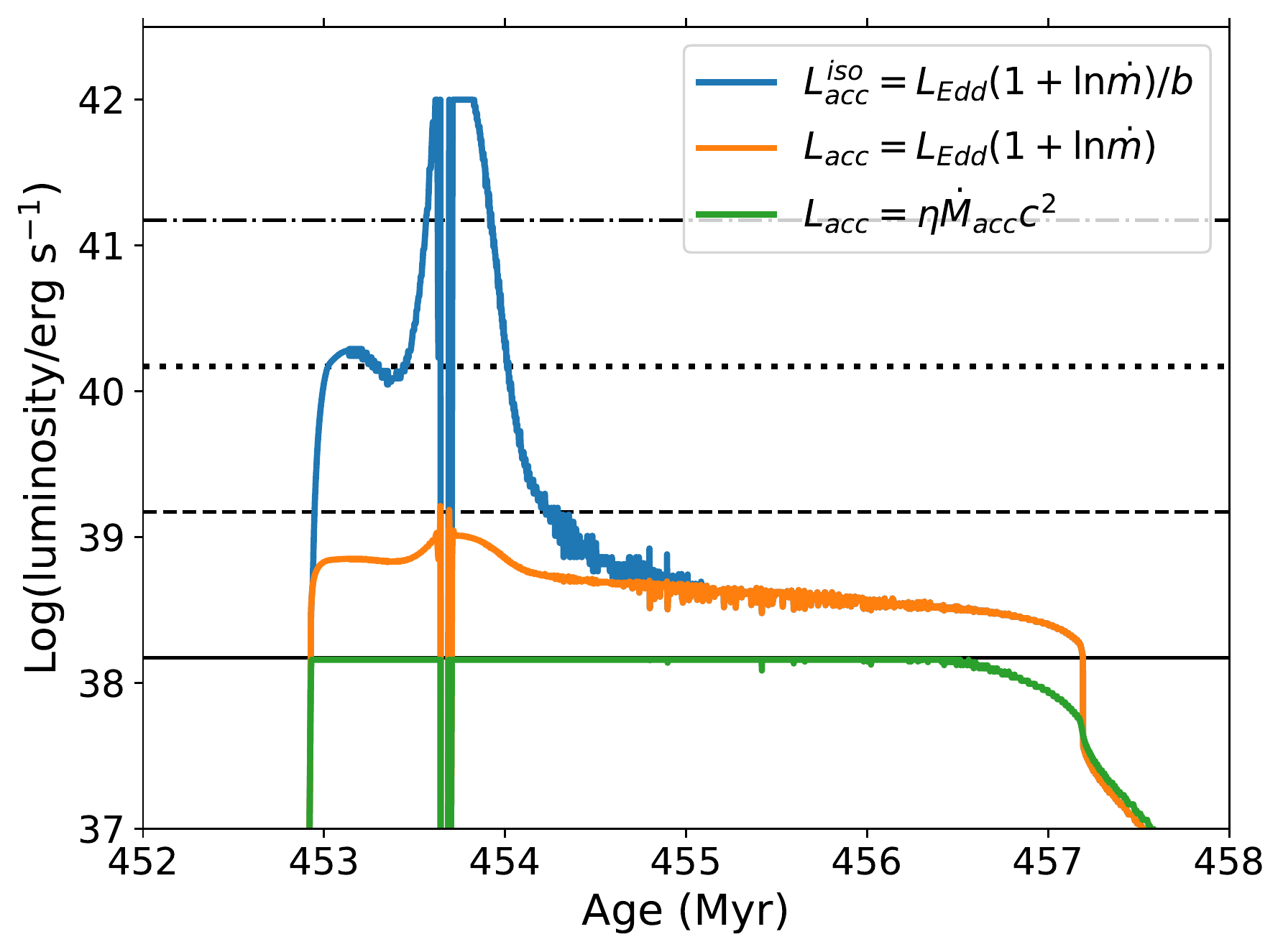}
        \caption{Estimated observed X-ray luminosity for a binary with M$^i_\mathrm{donor} =2.77$~M$_{\odot}$, M$^i_\mathrm{acc} =1.30$~M$_{\odot}$, and P$^i_\mathrm{orb} =5.38$~days under different assumptions. The solid green curve is the accretion luminosity corresponding to the mass-accretion rate onto the NS, assuming a fixed radiative efficiency ($\eta=0.1$). The solid orange curve is the accretion luminosity corresponding to a super-Eddington accretion disc following the \cite{1973A&A....24..337S} disc model given by Eq.~(\ref{edd_eq}). The solid blue curve is the isotropic-equivalent accretion luminosity corresponding to beamed super-Eddington emission following the \cite{2001ApJ...552L.109K} geometric beaming model (limited at $10^{42}$~erg~s$^{-1}$) given by Eqs.~(\ref{new_edd}) and (\ref{eq:beaming}). Four reference luminosity values, corresponding to $L_{\rm Edd}$ (solid black line), $10\,L_{\rm Edd}$ (dashed black line),  $100\,L_{\rm Edd}$ (dotted black line), and $1000\,L_{\rm Edd}$ (dot-dashed black line), are also plotted. The initial properties of this binary are highlighted with a magenta star in Fig.~\ref{peak_acc}.}
        \label{mdot}
\end{figure}
 
Figure~\ref{mdot} shows an example of a stable system with M$^i_\mathrm{donor} =2.77$~M$_{\odot}$, M$^i_\mathrm{acc} =1.30$~M$_{\odot}$, P$^i_\mathrm{orb} =5.38$~days, where we demonstrate the estimated observed X-ray luminosity under different assumptions. The solid green curve is the accretion luminosity corresponding to the amount of mass accreted by the NS (i.e. capped at exactly the Eddington limit), assuming a radiative efficiency following Eq.~(\ref{eta}). The solid orange curve is the accretion luminosity corresponding to a super-Eddington accretion disc following Eq.~(\ref{edd_eq}), where although the Eddington limit is locally satisfied at every point in the disc, the integrated luminosity of the disc can exceed the Eddington limit by a small factor. Finally, the solid blue curve is the isotropic-equivalent accretion luminosity corresponding to beamed super-Eddington emission following Eq.~(\ref{new_edd}) where the estimated observed luminosity can exceed the Eddington limit by up to a few orders of magnitude. 
For comparison, we mark with horizontal lines the luminosity of ULXs corresponding to $L_{\rm Edd}$ (solid black line), $10\,L_{\rm Edd}$ (dashed black line),  $100\,L_{\rm Edd}$ (dotted black line), and $1000\,L_{\rm Edd}$ (dot-dashed black line). For simplicity, we use Eq.~(\ref{Eq:Edd}) which is fixed for an initial accretor mass, because the amount of mass accreted does not change the accretion luminosity significantly. We note that since the prescription used for beamed emission (Section~\ref{acc}) is an approximation, we fix an upper limit for the calculated isotropic-equivalent accretion luminosities at $10^{42}$~erg~s$^{-1}$ in order to avoid unphysically high values. In any case, the time that our binaries spend at such high luminosities combined with the inferred very small beaming factors make binaries on that phase effectively non-detectable. In the remainder of the paper, we use the beamed model (i.e. equivalent to the blue curve based on Eqs.~(\ref{new_edd}) and (\ref{eq:beaming})) as our estimate of the observed X-ray luminosity of the binaries.

\begin{figure*}[t]
\centering
    \vfill
        \includegraphics[width=1.0\textwidth]{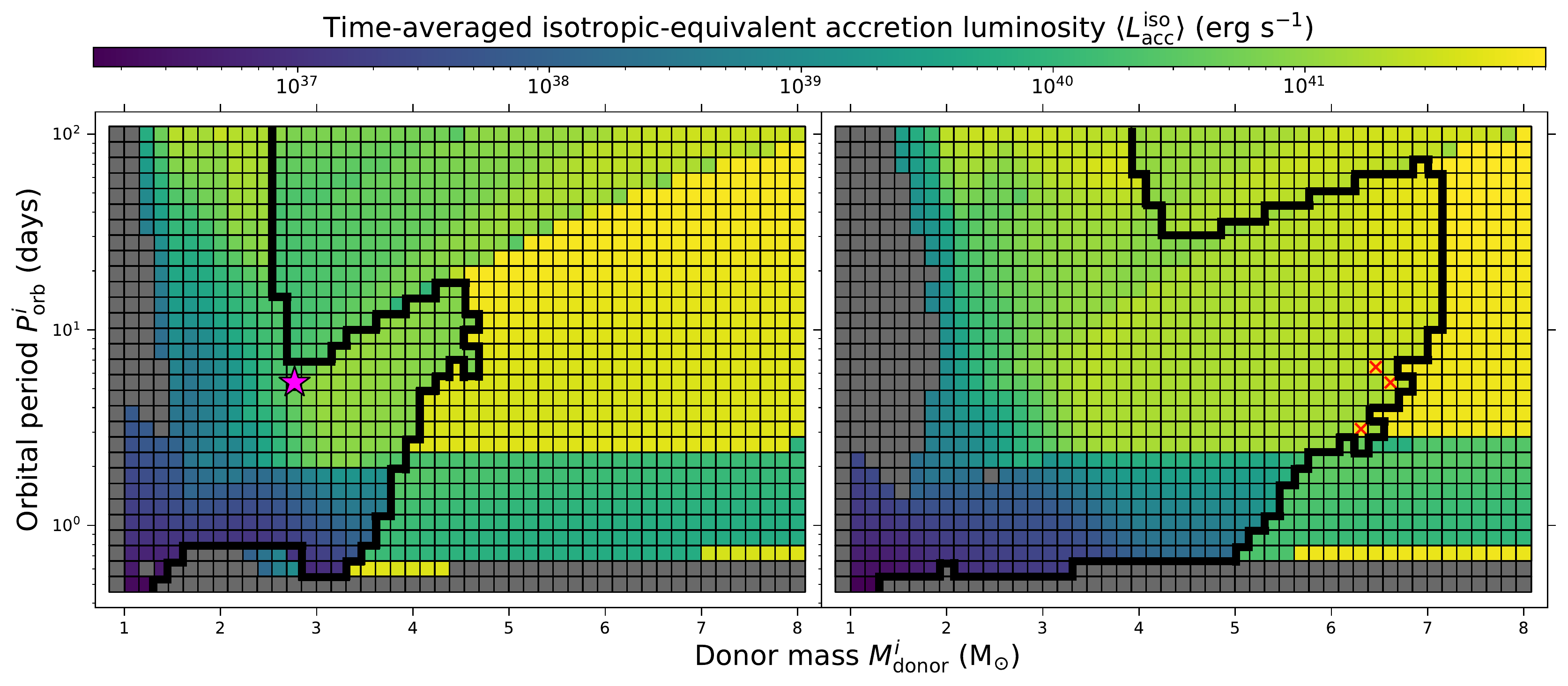}
        \caption{(Left) Time-averaged isotropic-equivalent accretion luminosities ($\langle L_{\rm acc}^{\rm iso}\rangle$) in LMXBs/IMXBs with a $1.3$~M$_\odot$ NS accretor that reached an instantaneous luminosity above $10\,L_{\rm Edd}$. These luminosities were calculated based on Eqs.~(\ref{new_edd}) and (\ref{eq:beaming}) for each binary and at each time-step, and averaged over the entire RLO lifetime. Grey colour denotes the sequences that never achieved an instantaneous luminosity above $10\,L_{\rm Edd}$. The systems that reach the highest accretion luminosities correspond to unstable systems, on the higher donor mass ends in both panels, before the onset of dynamical instability. The stable systems (enclosed by the solid black boundary) have a relatively low accretion luminosity  on average. The magenta star corresponds to the binary shown in Fig.~\ref{mdot}. (Right) Same as left but for LMXBs/IMXBs with $2.0$~M$_\odot$ NSs. The three red crosses correspond to systems that terminated due to numerical issues.}
        \label{peak_acc}
\end{figure*}

Figure~\ref{peak_acc} shows the time-averaged isotropic-equivalent accretion luminosities with respect to the initial parameters in both the grids. The magenta star is the binary shown in Fig.~\ref{mdot}. For accreting NSs of either $1.3\;$M$_\odot$ or $2.0\;$M$_\odot$, we find time-averaged isotropic-equivalent X-ray luminosities of $\langle L_{\rm acc}^{\rm iso}\rangle \simeq 10^{36}-10^{41}\;{\rm erg\,s}^{-1}$ (in some cases even up to $10^{42}\;{\rm erg\,s}^{-1}$).
These luminosities were calculated following Eq.~(\ref{new_edd}) and averaged over the entire RLO lifetime for systems that reached instantaneous luminosity above $10\;L_{\rm Edd}$. The stable systems are enclosed within the solid black boundary (for this and all subsequent figures). Comparing Fig.~\ref{rlo_cases1} with Fig.~\ref{peak_acc}, even the systems that undergo dynamical instability (L$_2$ overflow) are included. This is because, when using detailed binary evolution calculation, we are able to resolve the onset of the dynamical instability which is not instantaneous and is most often preceded by a short but intense phase of mass transfer. In fact, the systems that reach the highest luminosities correspond to the unstable systems on the higher donor-mass end in both panels of Fig.~\ref{peak_acc}. Therefore, binaries that would suffer dynamical instabilities and coalesce (likely producing Thorne-\.Zytkow objects hypothesised by \cite{1977ApJ...212..832T}) could also be observed as NS ULXs earlier in their evolution. However, the lifetimes of these binaries as ULXs is very short, and so this would act against their detectability.

\begin{figure*}[t]
    \vfill
    \centering
        \includegraphics[width=\linewidth]{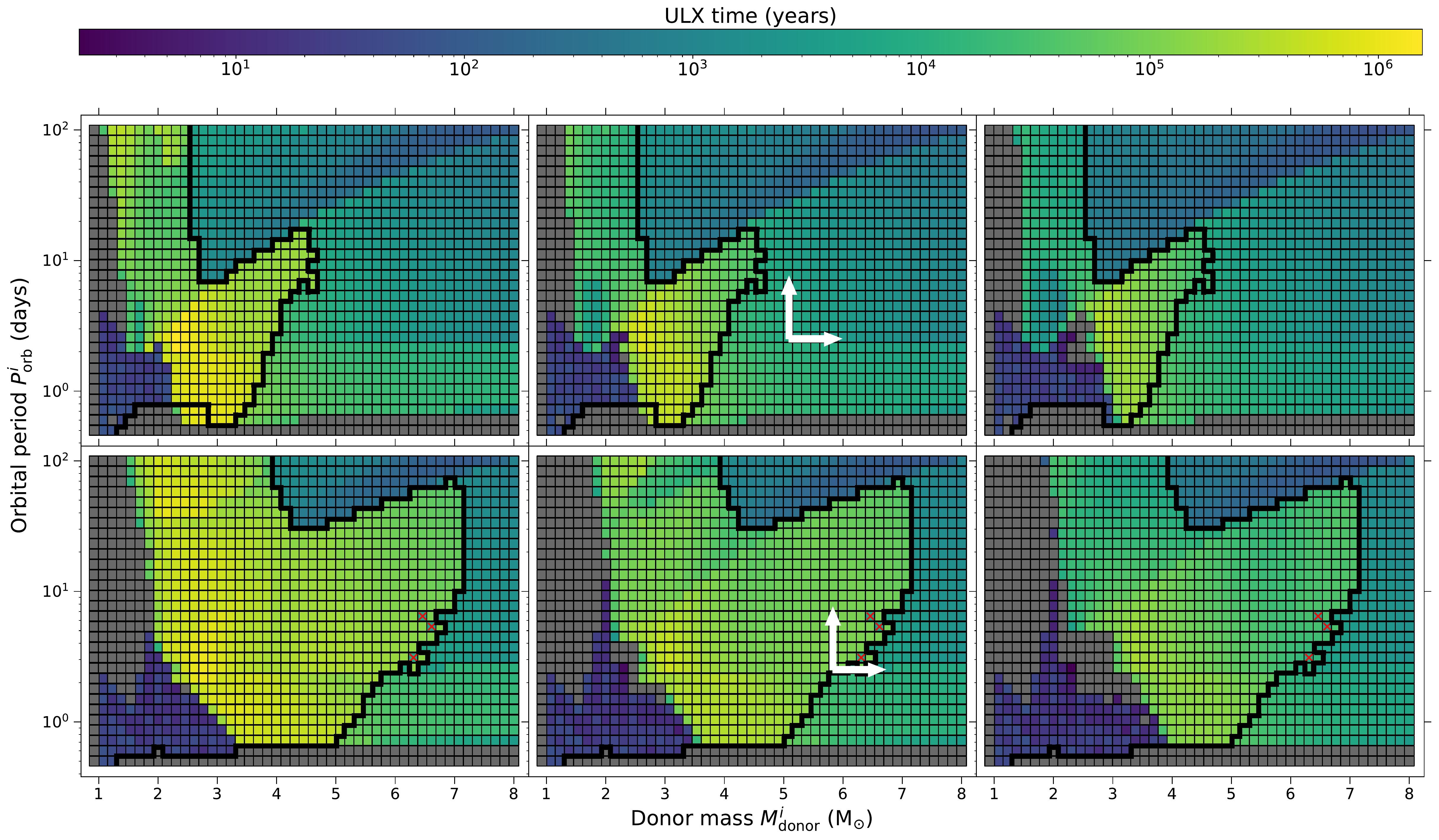}
        \caption{(Top row) ULX lifetime for LMXB/IMXB systems with a NS accretor mass of $1.3$~M$_{\odot}$. Left, middle, and right panels show the time that systems spent with $L_{\rm acc}^{\rm iso}$ above $10$, $100$, and $1000\,L_{\rm Edd}$, respectively. White arrows (middle panels) enclose the potential properties of M82~X-2 at the onset of RLO (see Sections~\ref{m82x2} and \ref{m82x2b}). (Bottom row) Same as the top row but for LMXB/IMXB systems with a NS accretor mass of $1.3$~M$_{\odot}$.}
        \label{result_1}
\end{figure*}

We define three X-ray luminosity ranges  >$10\,L_{\rm Edd}$, >$100\,L_{\rm Edd}$, and >$1000\,L_{\rm Edd}$, and calculate how long each system spends in each luminosity range. We refer to this time duration as ULX lifetime. The defined luminosity ranges are based on the observed pulsating ULX luminosities, which are in the range of $10$--$1000\, L_{\rm Edd}$ (Section \ref{sec:observations}). We compare the isotropic-equivalent accretion luminosities with these luminosity ranges. Figure~\ref{result_1} presents these results for LMXBs/IMXBs with $1.3$~M$_{\odot}$ and $2.0$~M$_{\odot}$ accretors. In systems with $1.3$~M$_{\odot}$ NSs, the longest ULX lifetime is $1.6\times 10^6$~years, corresponding to $M^i_\mathrm{donor} = 2.3$~M$_\odot$ and $P^i_{\mathrm{orb}}=2.16$~days, for observed luminosities of >$10\,L_{\rm Edd}$. In systems with $2.0$~M$_{\odot}$ NSs, the longest ULX lifetime is $1.1\times 10^6$~years, corresponding to $M^i_\mathrm{donor}=2.77$~M$_\odot$ and $P^i_{\mathrm{orb}}= 2.16$~days, again for observed luminosities of >$10\,L_{\rm Edd}$. The upper limits to ULX lifetimes are comparable to the ULX age estimate of $\sim~1$~Myr for NGC~1313~X-2 by \cite{2008AIPC.1010..303P} (Section~\ref{1313}). Looking at similar initial donor masses and initial orbital periods in both sets of LMXBs/IMXBs (going from top to bottom row in Fig.~\ref{result_1}), higher accretor mass corresponds to a much higher ULX lifetime as the stability area increases. As an example, for $M^i_\mathrm{donor}=5.0$~M$_\odot$ and $P^i_{\mathrm{orb}}=1.0$~days, the ULX time increases from $3.0\times 10^4$~years to $3.0\times 10^5$~years (going from lower to higher NS accretor mass).

In Fig.~\ref{result_1} (top row), from the leftmost panel to the rightmost (from >$10\,L_{\rm Edd}$ to >$1000\,L_{\rm Edd}$) the systems which initially have long ULX lifetimes ($M^i_\mathrm{donor}\sim 2.3$~M$_\odot$ and $P^i_{\mathrm{orb}}\sim 2.0$~days) either no longer appear on the plot or have a smaller ULX lifetime. Their isotropic-equivalent luminosities barely reach the higher cutoff values. Similarly, in the bottom row of Fig.~\ref{result_1}, most LMXBs/IMXBs with long ULX lifetimes ($M^i_\mathrm{donor}\sim 2.77$~M$_\odot$ and $P^i_{\mathrm{orb}}\sim 2.0$~days) for >$10\,L_{\rm Edd}$, do not reach luminosities >$1000\,L_{\rm Edd}$. This implies that binaries that are in the ULX phase for the longest time do not always achieve the highest luminosities. This effect is not seen in the unstable systems, which maintain their short ULX lifetime across the different ULX criteria. There is a part of the stable parameter space where the ULX lifetime decreases significantly, the ULX lifetime going from about $10^5$~years for systems outside this region to $10^1$--$10^2$~years (the boundary corresponding to $M^{i}_{\rm donor}<3.0$~M$_{odot}$ and $P ^{i}_{\rm orb}<2.0$). Their accretion luminosity is sub-Eddington for almost the entire mass-transfer phase. After a mass-transfer episode, these systems are left with a He core and thin H envelope which expands briefly during H-shell burning, which is enough to increase their luminosity to exceed the Eddington limit albeit for a very short period of time.

Most NS ULXs observed so far have been in the higher luminosity range ($100$--$1000\;L_{\rm Edd}$) potentially due to selection effects as more luminous ULXs have a higher chance of being observed. On the other hand, the short lifespans of NS ULXs compared to the age of the universe ($\sim 1.4\times 10^{10}$~yr) implies that it is unlikely to observe these systems in large numbers, which is consistent with the existing small NS ULX sample. The extra physical argument that the emission is highly beamed, especially for very luminous ULXs, further decreases the chances of detecting a large number of NS ULXs.

\begin{figure*}
    \centering
    \includegraphics[width=\linewidth]{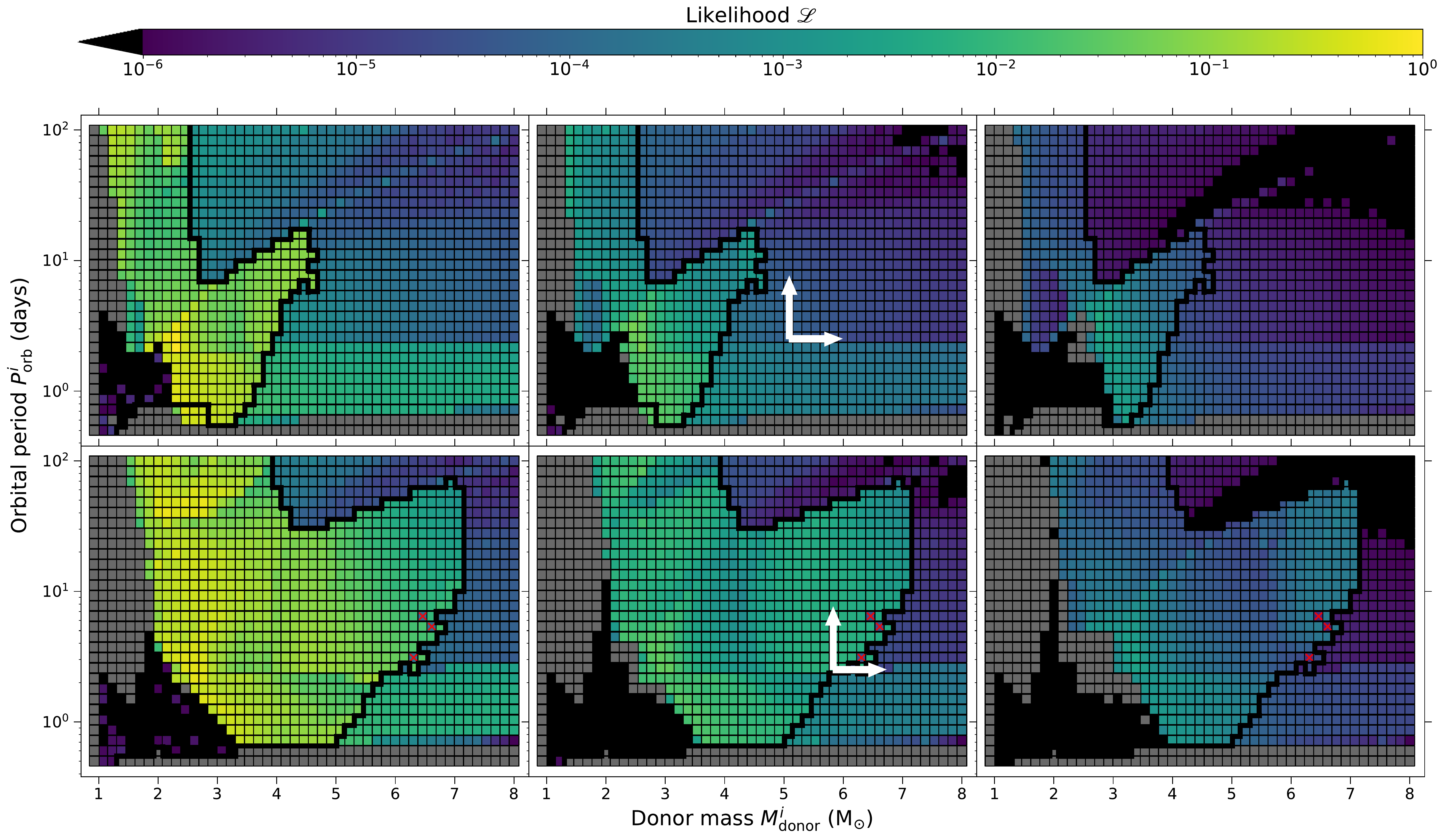}
    \caption{(Top row) Relative likelihood ($\mathscr{L}$) to observe a system as a ULX, as described by Eq.~(\ref{beam_eq}), for LMXBs/IMXBs with a $1.3$~M$_\odot$ NS. The panels are arranged as in Fig.~\ref{result_1}. Going from the leftmost to the rightmost panel  ($>10$ to >$1000\,L_{\rm Edd}$), $\mathscr{L}$ is higher for stable binaries with lower luminosities as their emission is not as strongly beamed. There is in addition some effect by the transition from case~A to case~B RLO. All values below $10^{-6}$ are shown in black as they correspond to insignificant likelihood. (Bottom row) Same as the top row but for LMXBs/IMXBs with a $2.0$~M$_\odot$ NS.}
    \label{likelihood}
\end{figure*}

As alluded to above, the probability of observing any of these LMXB/IMXB sources as ULXs depends on the beaming factor, the ULX lifetime, and the probability of the particular system being formed. We evaluate the first two factors in the following equation describing the likelihood of a ULX observation, which is the beaming factor integrated over the three defined ULX lifetimes:
\begin{equation}\label{beam_eq}
    \mathscr{L} = \int_{\mathrm{ULX}} b(t) dt.
\end{equation}
Calculating the likelihood using this equation for both the grids and for all three ULX lifetime criteria, we obtain a measure of the relative chance of observing each system in Fig.~\ref{likelihood} for the $1.3$~M$_\odot$ NS (top row) and the $2.0$~M$_\odot$ NS (bottom row) grids. All values below $10^{-6}$ are shown in black as they correspond to insignificant likelihood. The highest likelihood is for the stable systems with lower luminosities ($10\;L_{\rm Edd}$), where the mass-transfer rate never increases to extreme values and thus collimation is very small to almost negligible. Figure~\ref{likelihood} shows that the chance of a system being observed as a ULX decreases at higher luminosities. This is due to the fact that for higher luminosities the mass-accretion rates (and the mass-transfer rates) are extremely high which causes the emission from the system to be highly beamed. Also, the probability depends on the structure of the donor at the onset of RLO. Case A systems have lower mass-transfer rates and more isotropic emission and thus slightly higher probabilities of being observed than case~B. As mentioned before, there is a part in the stable parameter space where the ULX lifetime drops by many orders of magnitude (Fig.
~\ref{result_1}). Looking at the same systems in Fig.~\ref{likelihood}, the likelihood to observe them is negligible. These systems are super-Eddington for a very brief moment in time which is to the detriment of their detectability. Looking at Figs.~\ref{result_1} and \ref{likelihood}, it is evident that even though LMXBs are included in the initial parameter space, IMXBs (with donor masses $\gtrsim 2.0$~M$_{\odot}$) are better candidates for NS ULXs.

One caveat is that we assume that all systems have an equal probability of formation whereas in reality many systems might be formed at a higher rate than others. To account for these effects we would need to do population synthesis studies which would include exploring the formation probability of close NS + main sequence binaries and the distribution of their binary parameters. However, this is outside the scope of this study. We intend to study the formation rate of these systems in a future work.

Furthermore, one should always be careful when directly comparing our results to the observed population as our models give predictions about the whole population of ULXs with NS accretors, and not only the pulsating ones. The conditions for a NS to produce coherent pulses while also reaching super-Eddington luminosities are still unclear. If for example a strong magnetic field is required then perhaps the pulses are only observable at the beginning of the mass-transfer phase before any significant amount of material is accreted onto the NS, which could bury the magnetic field.

\begin{figure*}
\centering
\includegraphics[width=\linewidth]{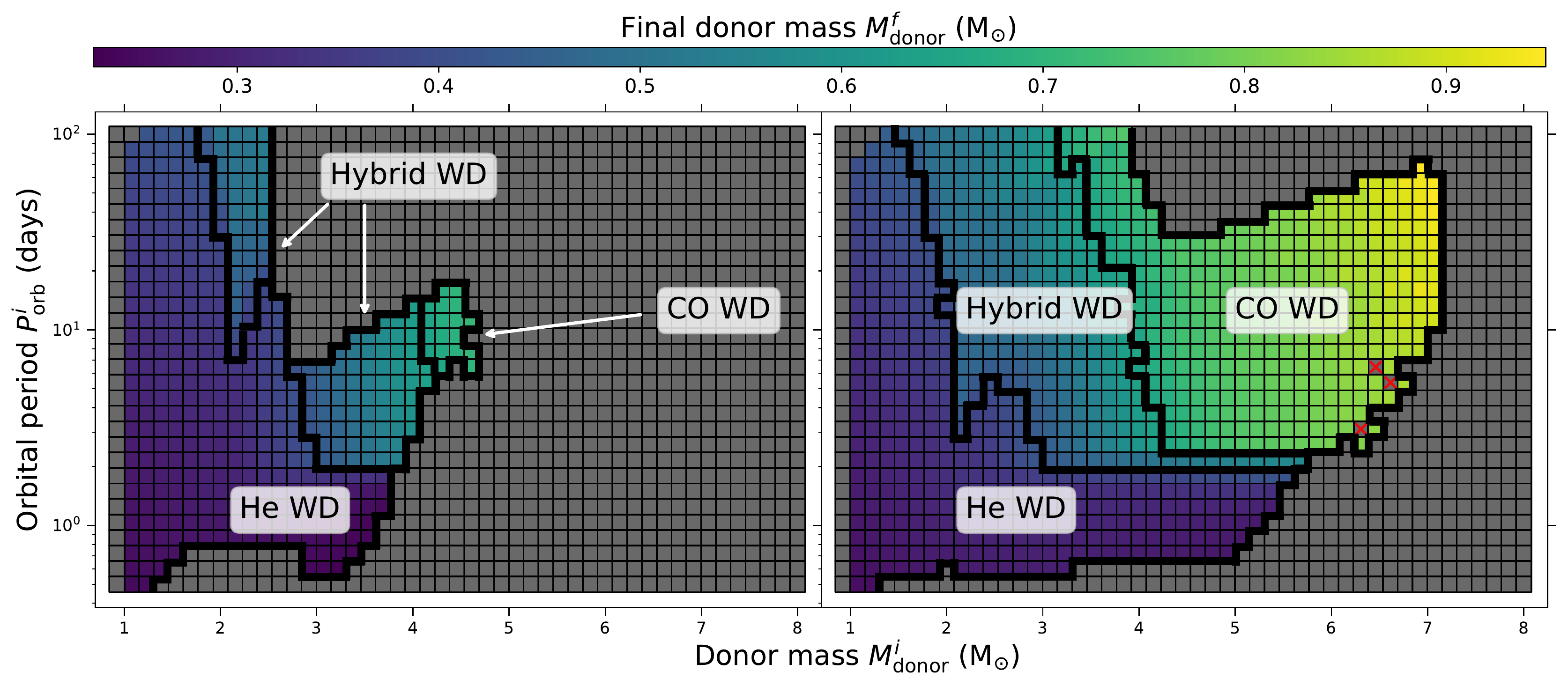}
\caption{(Left) Resulting distribution of final WD masses for LMXBs/IMXBs with $1.3$~M$_\odot$ NSs. The He~WDs are the least massive, followed by hybrid WDs, and the CO~WDs are the most massive. The outermost black boundary encloses the stable systems and the inner boundaries differentiate between the different WD types. (Right) Same as left but for LMXBs/IMXBs with $2.0$~M$_\odot$ NSs.}
\label{wd}
\end{figure*}

In cases where the mass-transfer sequence is stable throughout the binary evolution, we expect a NS--WD binary to be formed. Figure~\ref{wd} shows the type of white dwarf formed at the end and the final white dwarf masses for the $1.3$~M$_\odot$ NS (left panel) and the $2.0$~M$_\odot$ NS (right panel) grids. The superscript $f$ stands for final values. Overall the white dwarfs resulting from the $1.3$~M$_{\odot}$ NS grid are in the mass range of $0.23$--$0.71$~M$_{\odot}$. For systems with $2.0$~M$_\odot$ NS accretors,  the white dwarfs have an overall mass range of $0.25$--$0.95$~M$_{\odot}$. We used final total mass fractions of carbon to distinguish between different WD types. We define WDs with $>95\%$ carbon mass fraction as CO white dwarfs, $0.01$--$95\%$ as hybrid white dwarfs, and $<0.01\%$ as He white dwarfs. The initially higher donor masses and longer orbital periods result in a degenerate CO core with negligible helium on the surface. For some donor masses and orbital periods, the donor forms a degenerate CO core with a relatively large helium-rich envelope leading to hybrid WDs. In systems with low donor masses and short orbital periods, the donors end up as helium WD systems. The reason for the final fate of the latter class of systems is that  due to deep envelope stripping early in the evolution of the donor, combined in some cases with the low initial mass of the donor, the final stripped helium cores are not massive enough to ignite helium.

\begin{figure*}
\centering
\includegraphics[width=\linewidth]{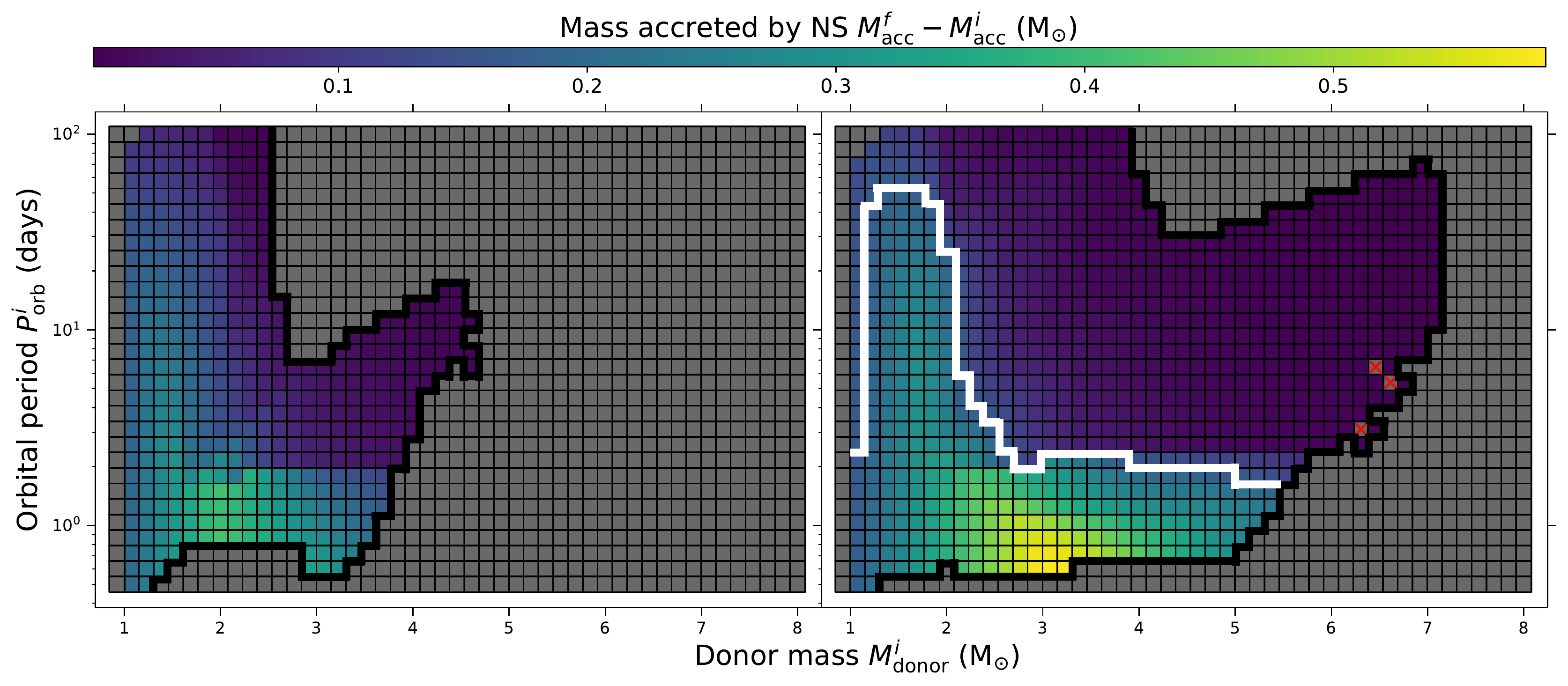}
\caption{(Left) Mass accreted by the NS ($M_{\rm acc}^{f}-M_{\rm acc}^I$) for LMXBs/IMXBs with a $1.3$~M$_\odot$ NS and $\beta = \mathrm{max}[0.7, 1-{\dot{M}_{\mathrm{Edd}}}/{\dot{M}_{\mathrm{donor}}}]$. Binaries producing the lowest WD masses show the highest accretion on the NS. (Right) Same as left but for LMXBs/IMXBs with a $2.0$~M$_\odot$ NS. The systems below the white boundary are those where the accretor has accreted enough mass to exceed the maximum mass limit for a NS, here assumed to be $2.17\;M_\odot$.}
\label{ns}
\end{figure*}

Figure~\ref{ns} shows the mass accreted by the NSs, for systems with $1.3$~M$_\odot$ (left) and $2.0$~M$_{\odot}$ NS (right) accretors. We should reiterate here our assumption of non-conservative mass transfer with $\beta = \mathrm{max}[0.7, 1-{\dot{M}_{\mathrm{Edd}}}/{\dot{M}_{\mathrm{donor}}}]$ and that the accretion by the NS is Eddington limited. The $1.3$~M$_\odot$ NSs end up accreting $0.004$--$0.415$~M$_{\odot}$ of mass by the end of the mass-transfer phase, the systems with values around $M_{\mathrm{donor}}^i = 2.0$~M$_\odot$ and $P_{\rm{orb}}^i = 0.87$~day accreting the most. The amount of mass accreted is not enough to collapse the NS. The $2.0$~M$_{\odot}$ NSs accrete mass in the range of $0.002$--$0.585$~M$_{\odot}$, with the systems with values around $M_{\mathrm{donor}}^i = 3.0$~M$_\odot$ and $P_{\rm{orb}}^i = 3.1$~days accreting the highest amount. We compare our results to the maximum NS mass known. The most massive precisely measured NS is PSR~J0348+0432 with a mass of $2.01\pm0.04\;M_\odot$ \citep{2013Sci...340..448A}. However, recently, a candidate with a higher NS mass of $2.17^{+0.11}_{-0.10}\;M_\odot$ was announced \citep[PSR~J0740+6620;][]{2019arXiv190406759C}. Although the error bar of the latter source is relatively large, we take $2.17\;M_\odot$ as our assumed upper limit. This value is also supported by constraints on GW170817 based on combined gravitational wave and electromagnetic observations \citep{2017ApJ...850L..19M}.

Looking at the final NS masses in Fig.~\ref{ns} (right panel), we see a certain population enclosed by a white boundary. In these systems the accretor has accreted enough material to overcome the neutron degeneracy pressure that is supporting the star against gravitational collapse (assuming an upper NS mass limit of $2.17\;M_\odot$). Therefore, according to the maximum NS mass limit assumed, these NSs will collapse to form BHs. 

Net accretion on the NS can be broadly described by,
\begin{equation}
    \Delta M_{\mathrm{acc}} = \langle\dot{M}_{\mathrm{acc}}\rangle \times \Delta t_{\mathrm{X}},
\end{equation}
where $\Delta t_{\mathrm{X}}$ is the lifetime as an X-ray binary (or the overall mass-transfer phase) and $\langle\dot{M}_{\mathrm{acc}}\rangle$ is the average accretion rate onto the NS. A high amount of accreted mass is achieved with a combination of both a large accretion rate and a long time duration over which it occurs. 
For LMXBs/IMXBs that reach up to Eddington mass-transfer rates for only a small part of the mass-transfer phase, to zeroth order, $\Delta M_{\mathrm{acc}}\simeq 0.3\times \Delta M_{\rm donor}$. In contrast, for the brightest ULXs, where most of the mass-transfer happens at a highly super-Eddington rate, $\langle\dot{M}_{\mathrm{acc}}\rangle \simeq \dot{M}_{\rm Edd}$ and $\Delta M_{\mathrm{acc}} << \Delta M_{\rm donor}$. Even if the accretion onto the NS was allowed to reach a few times the Eddington limit, the net amount of accreted material by the NS in the brightest ULXs would still be small.

\begin{figure*}
\centering
\includegraphics[width=\linewidth]{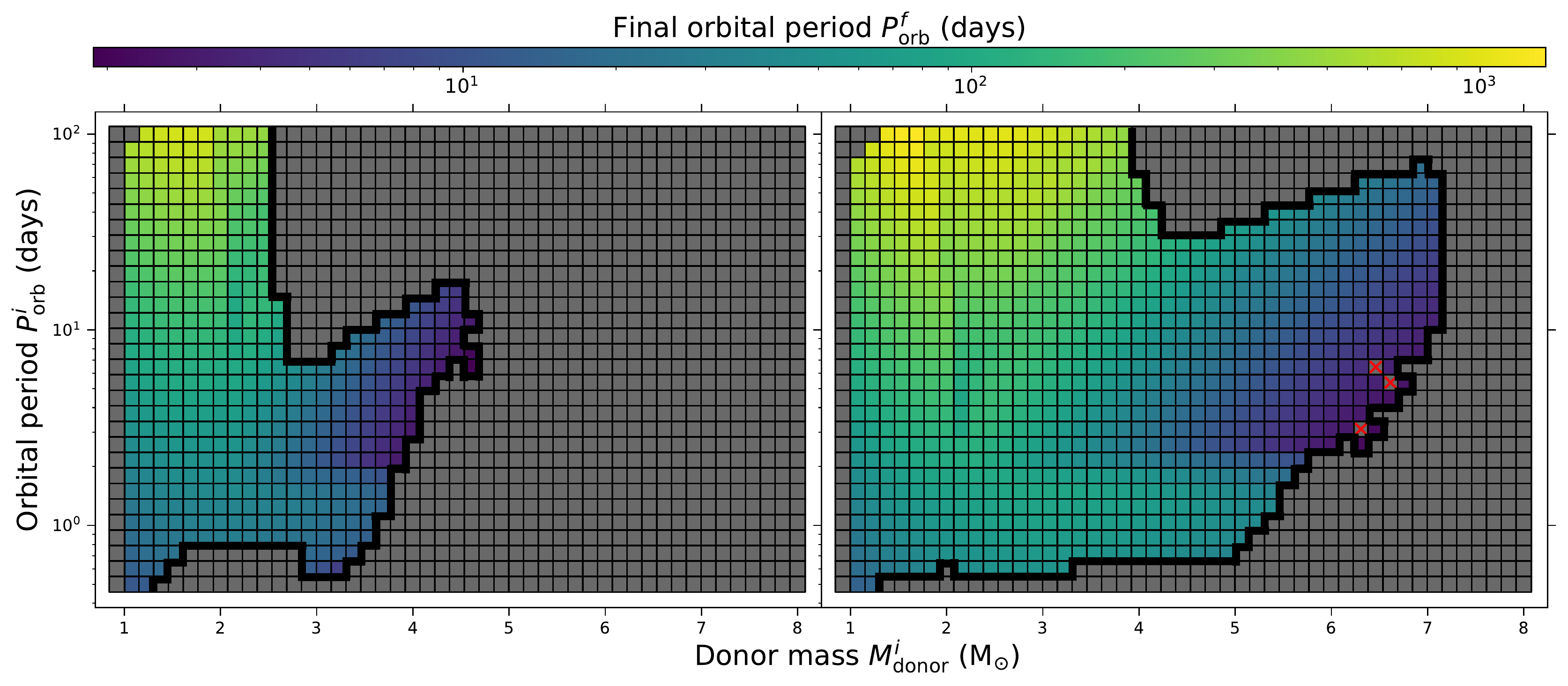}
\caption{(Left) Final orbital periods $(P^{f}_{\mathrm{orb}})$ for LMXBs/IMXBs with a $1.3$~M$_\odot$ NS. The longest final orbital periods result from initial periods $\gtrsim 10.0$~days and donors with masses less than about 2.75~M$_{\odot}$. The shortest final orbital periods are for systems that formed the heaviest WDs, originating from donors of $\gtrsim 3.5$~M$_{\odot}$. (Right) Same 
as left but for LMXBs/IMXBs with $2.0$~M$_\odot$ NSs. }
\label{pd}
\end{figure*}

The final orbits in most systems that underwent stable mass transfer are wide because during the mass-transfer phase, after the orbit shrinks initially, the donor evolves to become less massive than the accretor causing the system to widen significantly. These widened orbits can be seen in Fig.~\ref{pd} which shows that the final orbital periods can be as large as $5$--$7$ times the initial orbital periods. Some of these binaries may be observable by Gaia. \cite{2019ApJ...886...68A} postulate that Gaia can detect and measure the properties of hidden wide binaries with WD and NS components using astrometric observations. The comparison of NS--WD observations to final orbital parameters from our calculations might help in understanding which binaries have undergone a super-Eddington mass-transfer phase in the past and could have been observed as ULXs, as well as constrain binary evolution physics such as the accretion efficiency.

%%%%%%%%%%%%%%%%%%%%%%%%%%%%%%%%%%%%%%%%%%%%%%%%%%%%%%%%%%%%%%%%%%%%%%%%%%%%%

\section{Discussion}\label{sec:discussion}

\subsection{Comparison to observations of M82~X-2}\label{m82x2b}
Since M82~X-2 has fairly well constrained parameters (see Section~\ref{m82x2} and Table~\ref{tab:obs_data}), we can compare it to our results and see if our results help to explain the observations. We look at the middle panels in both rows in Figs.~\ref{result_1} and \ref{likelihood}. Since the currently observed donor is more massive than the accretor, the orbit will shrink as mass is lost following the orbital angular momentum balance equation as described in Eq.~(\ref{analytical}) for a non-conservative mass-transfer phase. Therefore, we use the observed parameters as lower limits of the initial binary configuration.

We only consider the luminosity range of $100\,L_{\rm Edd}$ because it is comparable to the observed luminosity of about $1.8\times 10^{40}$~erg~s$^{-1}$. \cite{2014Natur.514..202B} assumed an accretor of $1.4$~M$_{\odot}$ and estimated the donor to be $\gtrsim 5.2$~M$_{\odot}$. For a $1.3$~M$_{\odot}$ accretor mass, the donor mass is $\gtrsim 5.1$~M$_{\odot}$, using the same binary mass function. The initial parameter estimates are enclosed by the white arrows in Figs.~\ref{result_1} and \ref{likelihood} (middle panels in both rows). The potential initial systems have ULX lifetimes as long as $0.7\times 10^4$ years and a relative peak likelihood of $0.7\times 10^{-4}$. However, if we assume a higher accretor mass of $2.0$~M$_{\odot}$, the donor mass is $\gtrsim 5.83$~M$_{\odot}$. The longest ULX lifetime in this case for the potential initial systems is about $1.1\times 10^5$ years with a peak likelihood of $0.7\times 10^{-2}$. Taking these numbers at face value, it is clear that an initially heavy NS is preferred in order to explain the currently observed properties of M82~X-2. This cannot be excluded as the $1.4$~M$_{\odot}$ NS used by \cite{2014Natur.514..202B} was an assumption. We should note, however, that for both NS masses, the peak of the relative likelihood does not lie very close to the observed limits we have for the current properties of M82~X-2. This is not necessarily problematic, as in order to calculate the actual probability density distribution of what the properties of NS ULXs ought to be, based on our model, we need to multiply the relative likelihood shown in Fig.~\ref{likelihood} with the `prior' probability of forming a NS binary with these initial conditions. This convolution might significantly shift the peak of the resulting probability distribution. We leave this calculation for future work. \cite{2015ApJ...802L...5F}, who followed such an approach, estimated the most probable initial donor mass of any NS ULX to be $3.0$--$8.0$~M$_{\odot}$, and the initial orbital period to be $1.0$--$3.0$~days. This parameter space lies within our results.

\subsubsection{Comparison with high-mass X-ray binary ULXs}

With our work we aim to explore the possibility of an LMXB or IMXB origin for pulsating ULXs. However, there are at least three known NS ULXs which are HMXBs, namely NGC~7793~P13, M51~ULX-7, and more recently NGC 300 ULX1.

NGC~7793~P13 has a luminosity of about $100\;L_{\rm Edd}$. With the presence of an observed high-mass donor ($>18\;M_\odot$, Section~\ref{7793} and Table~\ref{tab:obs_data}), this system cannot be explained by our LMXB/IMXB models. In our NS ULXs, extreme mass ratios produce unstable binaries which reach the ULX luminosity observed by the source for a timescale of $\sim 10^2$--$10^4 $~years before initiating L$_2$ overflow. However, this system is a HMXB with a donor of mass $18.0$~M$_{\odot}<M_{\mathrm{donor}}<23.0$~M$_{\odot}$ and is outside the range of our simulations. 

M51~ULX-7 is another HMXB in a ULX phase and containing an accreting NS. \cite{2019arXiv190604791R} suggest that the binary consists of an OB giant star ($M_{\mathrm{donor}}\lesssim 8.0$--$13.0$~M$_{\odot}$) and a NS with a magnetic field of $10^{12}$--$10^{13}$~G. The most recent NS ULX observed to have a super-giant donor is NGC 300 ULX1, with a donor star mass of $M_{\mathrm{donor}}\gtrsim 8.0$~M$_{\odot}$ \citep{2019arXiv190902171H}. These two systems are also not satisfactorily explained by our calculations, since we find such systems have a ULX lifetime of $\lesssim10^4$~years (for donors of $8.0$~M$_{\odot}$).

\cite{2019arXiv190304995Q} explored the stability of super-giant X-ray binaries with a very high donor mass  compared to accretor mass (donors up to 20 times more massive than the accretors), and found that the ULX phase could be long lasting ($\sim 0.4\times 10^{6}$~years) for such an extreme mass ratio if the super-giant star has a H/He gradient in the layers beneath its surface. These latter authors demonstrated that such systems can evolve on a nuclear timescale with a BH or NS accretor, even for binaries where the donor mass is up to 20 times the accretor mass. In their binary evolution models, the donor stars rapidly decrease their thermal equilibrium radius and can therefore cope with the inevitably strong orbital contraction imposed by such a high mass ratio.
These binaries could be post-CE systems, where the super-giant donor star has lost most of its hydrogen-rich envelope and the remaining one is enriched in helium. Recent 1D hydrodynamical simulations of a CE phase between a super-giant donor and a NS predict the formation of such binary configurations \citep{2019ApJ...883L..45F}.

\subsection{Effect of spin-orbit coupling on orbital evolution}\label{ls-coup}

As we already described in Section~\ref{sync_time}, our binary evolution calculations take into account the effect of spin-orbit coupling by modelling the tidal interactions, internal rotation, and mass transfer and/or loss. Here we want to quantify the effect of spin--orbit coupling on the orbital evolution and hence on the stability of mass transfer. We perform tests to calculate the orbital evolution in some limiting cases  and compare our numerical calculations with others available in the literature, as well as analytical solutions.

For simplicity, we assume a fully non-conservative mass transfer ($\beta = 1.0$) in the following calculations, i.e. no mass is accreted by the NS. Also, we assume that the binary remains always in synchronous rotation with its components so that the spin period of the donor is the same as the orbital period. This assumption corresponds to an infinitely efficient spin--orbit coupling. Under these assumptions, we obtain the analytical solution for the orbital evolution as follows (see Appendix~\ref{append1} for the detailed derivation):
\begin{align}\label{analytical}
    \nonumber \frac{\dot{a}}{a} = \bigg[\frac{M_{\mathrm{donor}} M_{\mathrm{acc}} a^{2}}{M_{\mathrm{donor}} M_{\mathrm{acc}} a^{2} - 3 I_{\mathrm{donor}} M}\bigg]\times \bigg[\dot{M}_{\mathrm{donor}} \bigg(\frac{2M_{\mathrm{donor}}}{M_{\mathrm{acc}}M} - \\ \frac{2}{M_{\mathrm{donor}}} +
    \frac{1}{M}\bigg) - \bigg(\frac{2\dot{I}_\mathrm{donor}M + I_\mathrm{donor}\dot{M}_{\mathrm{donor}}}{M_{\mathrm{donor}} M_{\mathrm{acc}} a^{2}}\bigg)\bigg],
\end{align}
where $I_{\rm donor}$ is the moment of inertia of the donor.
When the system loses mass from the vicinity of the accretor, according to Eq.~(\ref{analytical}) the orbit shrinks. This decrease occurs as long as $M_{\mathrm{donor}}\gtrsim 1.28\;M_{\mathrm{acc}}$ \citep{2001nsbh.conf..337T}. The orbit expands if mass is being lost from the approximate vicinity of the more massive star.

\begin{figure}
\centering
\includegraphics[width=\linewidth]{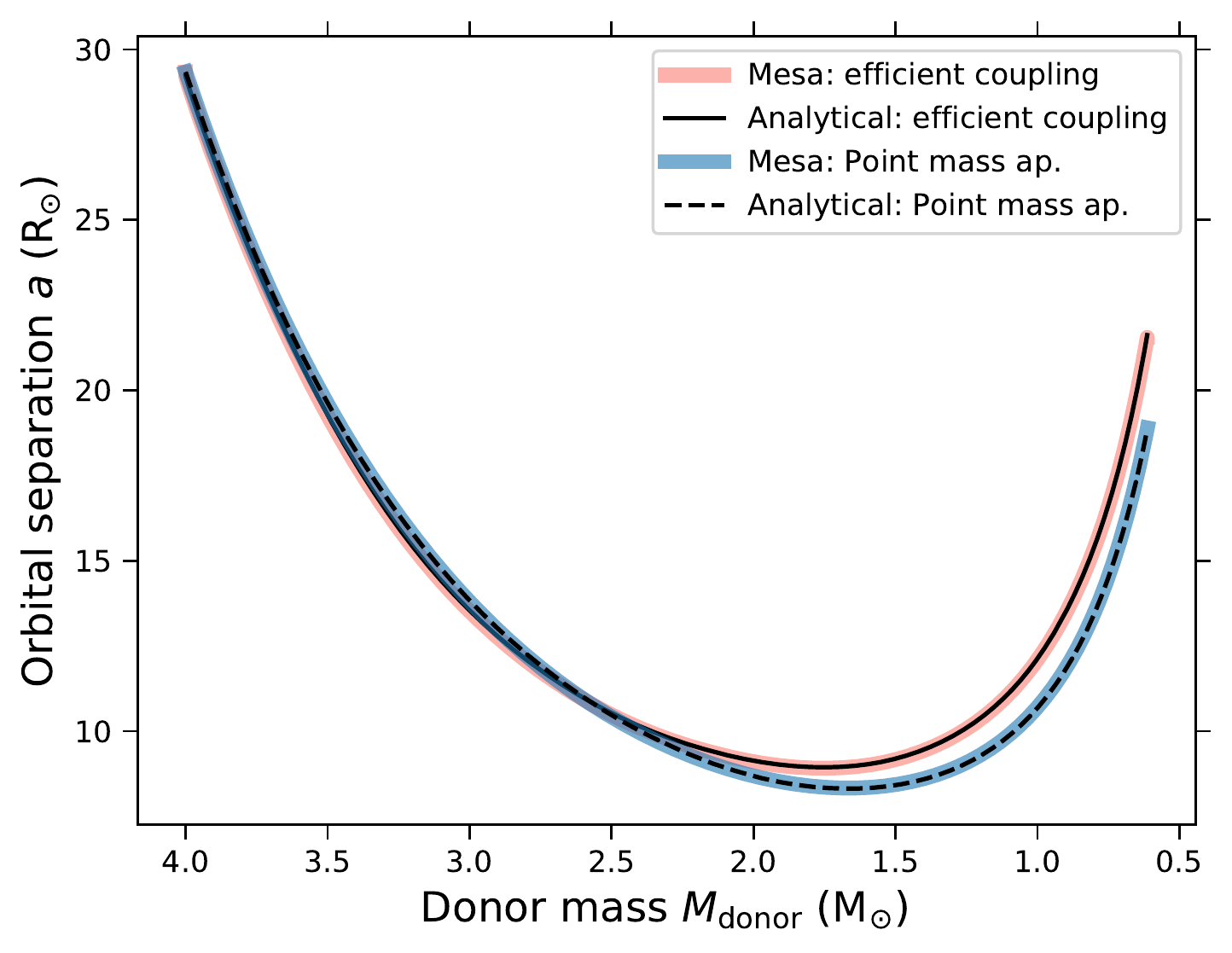}
\caption{Binary separation evolutionary sequences for a system with initially $M_{\mathrm{donor}}=4$~M$_{\odot}$, $M_{\mathrm{acc}}=1.3$~M$_{\odot}$ and $8$~days orbital period ($\beta = 1.0$). Different cases showing spin--orbit and no spin--orbit coupling are shown in comparison with the analytical equation calculations (described by Eq.~(\ref{analytical})).}
\label{test1}
\end{figure}

We run models using MESA, including angular momentum losses from spin--orbit coupling and mass lost from the system, while ignoring gravitational wave radiation and magnetic braking. The results of this test are shown in Fig.~\ref{test1} (blue and orange curves), and they are compared to the analytical solution of Eq.~(\ref{analytical}) (solid and dashed black curves), where $I_\mathrm{donor}$ and $\dot{I}_\mathrm{donor}$ are taken from numerical stellar structure calculations. For two extreme cases of spin--orbit coupling, we first consider the case where the donor star is an extended body ($I_\mathrm{donor}\neq 0$) and there is efficient coupling of donor spin and orbit and thus an efficient transfer of angular momentum between spin and orbit (orange curve in the figure). This keeps the spin of the donor synchronised with the orbit throughout the binary evolution and the binary is `tidally locked'. The second case is where both binary components are approximated as point masses and there is no exchange of angular momentum between the donor (blue curve in the figure) and the orbit ($I_\mathrm{donor}=0$). The two cases described above are extremes of spin--orbit coupling; any realistic solution would be in-between them. The MESA calculations are consistent with an analytic solution and the coupling does have a slight effect on the orbital evolution.

\begin{figure}
\centering
\includegraphics[width=\linewidth]{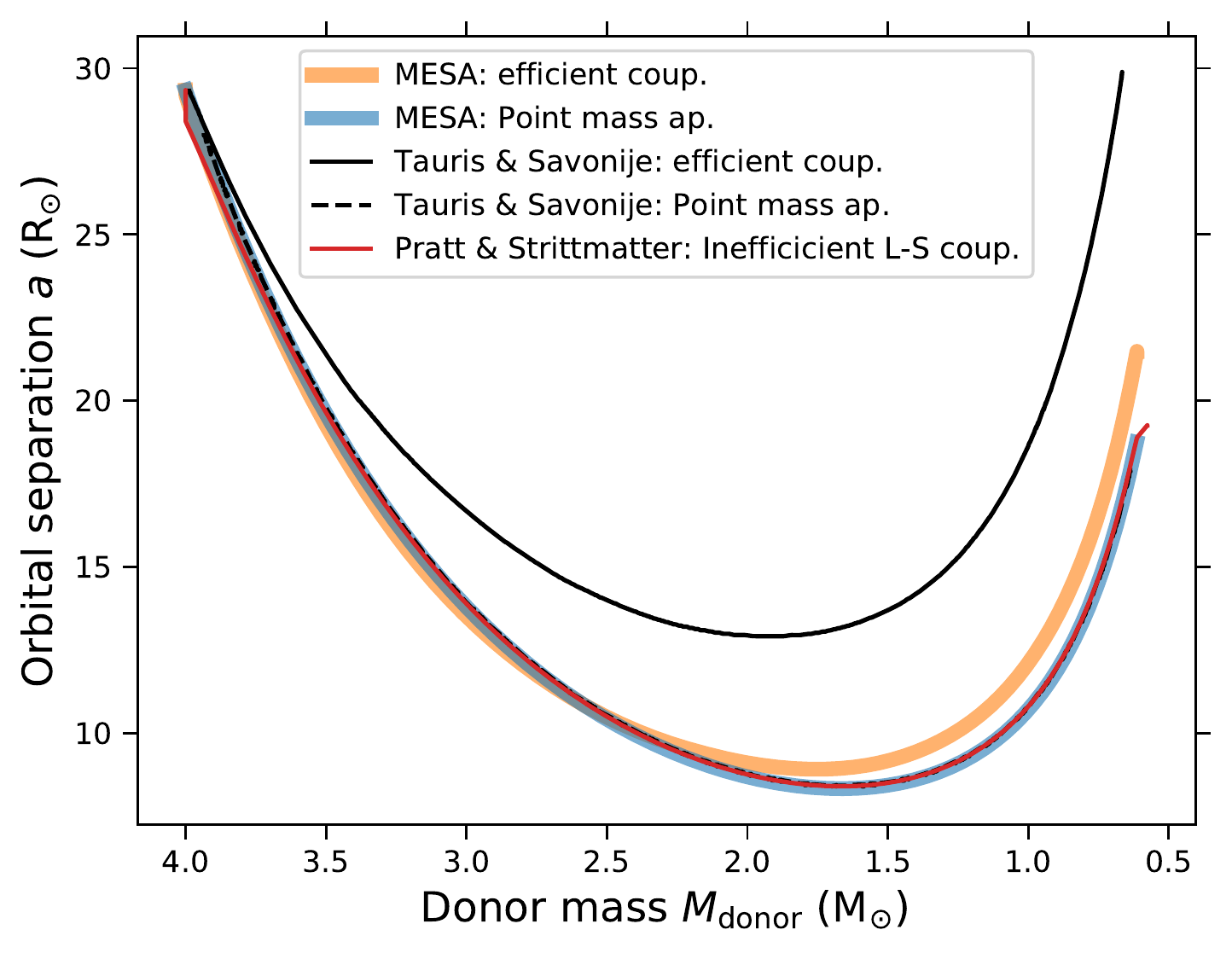}
\caption{Binary separation evolutionary sequences for a system with initially $M_{\mathrm{donor}}=4$~M$_{\odot}$, $M_{\mathrm{acc}}=1.3$~M$_{\odot}$ and $8$~days orbital period ($\beta =  {\it max}~\{0, 1-\dot{M}_{\mathrm{Edd}}/\dot{M}_{\mathrm{donor}}\}$). Different cases showing efficient spin--orbit and point mass approximation are shown in comparison with similar tracks from \cite{2001nsbh.conf..337T} (Fig.~2 in their paper).}
\label{test2}
\end{figure}

We also compare two MESA runs with the results of \cite{2001nsbh.conf..337T} (their Fig.~2). In this comparison we assume that $\beta = \mathrm{max}[0, 1-{\dot{M}_{\mathrm{Edd}}}/{\dot{M}_{\mathrm{donor}}}]$, and therefore that mass is transferred conservatively until the Eddington limit is reached. Furthermore, gravitational wave radiation and magnetic braking are included. This comparison is shown in Fig.~\ref{test2}. We again consider two cases, one with a point mass approximation (blue and dashed black curves) and the other one with efficient spin-orbit coupling (orange and solid black curves). While the evolutionary sequences for the point mass approximation (blue curve) match those from \citet{2001nsbh.conf..337T} (dashed black curve), the sequences corresponding to an efficient spin--orbit coupling show a discrepancy. The effect of angular momentum transport between the donor star and the orbit seems to be much larger  in \citet{2001nsbh.conf..337T} (solid black curve) compared to our calculated MESA tracks and analytic solutions (orange curve). In an attempt to understand this discrepancy, we also considered the case of highly inefficient synchronisation during RLO. In this scenario, during the high mass-loss phase, the donor loses mass (and angular momentum) which causes its spin to decrease. Since the synchronisation is not efficient, the orbit does not regulate the spin of the star as quickly as angular momentum is being lost from the star. \cite{1976ApJ...204L..29P} explore this case as a means to have a wider orbit and we also present this run in Fig.~\ref{test2} as the solid red line. This model has highly efficient spin--orbit coupling until the point the donor almost overflows its Roche lobe and begins mass transfer after which the spin--orbit coupling is highly inefficient. However, this reasoning does not seem to explain the described discrepancy either. Unfortunately, it is impossible to further investigate this apparent discrepancy, as the original calculations and code setup from \citet{2001nsbh.conf..337T} were not available.

In conclusion, due to the fact that the widening effect of spin--orbit coupling is weak in our calculations, we infer that spin--orbit coupling can affect the mass-transfer stability but is a second-order effect.

\section{Conclusions}\label{sec:conclusion}

There is much intrigue surrounding the nature of ULXs because they challenge our understanding of accretion onto compact objects at high rates and of the evolution of binaries containing compact objects. After the discovery of the first pulsating ULX (M82~X-2), NS~X-ray binary systems were known to constitute a (possibly major) sub-population of ULXs. 

NS IMXBs, with donor masses of $2.0-10.0$~M$_{\odot}$, have been proposed in the past as possible ULXs \citep{2000ApJ...530L..93T, 2015ApJ...802L...5F}. Their extreme mass ratio leads to intense mass transfer, which can become  highly super-Eddington. Due to changes in the structure of the accretion disc as the mass-transfer rate exceeds the Eddington limit,  face-on observers can infer super-Eddington isotropic-equivalent accretion luminosities, even if the accretion onto the NS is actually limited to the Eddington rate. This is because the increased radiation pressure causes the inner parts of the disc to puff up, creating a funnel, which in turn can cause beaming of the outgoing radiation. These effects can increase the observed isotropic-equivalent luminosities of some accreting NSs to a few orders of magnitude above the Eddington limit. However, the higher the accretion luminosity the lower the chances of actually observing the source since the emission is highly beamed and depends on the angle of viewing. 

In this work, we attempt to explain super-Eddington luminosities observed in NS~ULXs by calculating extensive grids of binary evolution models of LMXBs/IMXBs with donor stars in the mass range of $0.92$--$8.0\;$M$_\odot$ using the MESA code. We specifically investigate the stability and rate of mass transfer, taking into account rotation and tidal effects, as well as the observable properties and the final outcomes of these binaries. The results we obtain can be summarised as follows:

\begin{itemize}
   \item It has been demonstrated previously that the stability of mass transfer in a binary system primarily depends on the structure of the envelope of the donor star at the onset of RLO. From our numerical modelling, including stellar rotation and tides, we find that the most extreme mass ratios ($q=M_{\rm acc}/M_{\rm donor}$) that are able to evolve through long-term stable mass transfer are $0.5$ for donors with convective envelopes and 0.28 for donors with radiative envelopes.
   \item For more extreme mass ratios, mass transfer via the first Lagrangian point (L$_1$) cannot keep the donor star contained within its Roche lobe, as the Roche-lobe radius is shrinking faster than the radius of the donor star. In these cases, the donor star expands well outside its Roche lobe, and its volume can even exceed the volume enclosed by the equipotential surface that passes from the second Lagrangian point (L$_2$). At this stage, mass loss from the L$_2$ point will occur, which causes significant loss of orbital angular momentum, thereby accelerating the shrinkage of the orbit and leading to the onset of a CE event. We have calculated L$_2$ lobe volumes and derived fitting formulae (accurate to errors less than 1\% for all values of $q$) to calculate the volume-equivalent radii of donor stars at the onset of mass loss via L$_2$. In our calculations, we self-consistently identify the onset of the L$_2$ overflow and consider this to be the beginning of an irreversible dynamical instability. 
   \item We demonstrate that LMXBs/IMXBs can produce NS-ULXs with typical time-averaged isotropic-equivalent X-ray luminosities between $10^{39}$ and $10^{41}\;{\rm erg\,s}^{-1}$ on a timescale of up to $\sim\!1.0\;{\rm Myr}$ for the lower luminosities. We present plots of their likelihood of detection, taking into account beaming effects. Based on these results, IMXBs are better candidates for NS ULXs than LMXBs.
   \item We find that the beamed super-Eddington accretion disc model provides the very high isotropic-equivalent X-ray luminosities that have been observed for NS ULXs. The model also explains the low detectability of these sources as highly beamed emission and short ULX lifetimes are common properties of the sources.
   \item LMXBs/IMXBs cannot explain the origin of all NS~ULXs given that some of their donor stars are massive (i.e. therefore HMXBs). However, we confirm that IMXBs, specifically, are strong candidates for a large fraction of NS~ULXs. For the most well-studied NS~ULX, M82~X-2, our LMXB/IMXB mass-transfer sequences with $2.0$~M$_{\odot}$ NSs better explain the observations compared to systems with $1.3$~M$_{\odot}$ NSs.
   \item We obtain three populations of WDs at the end of the stable mass-transfer phase: He~WDs, CO~WDs, and hybrid WDs which have a CO core and a significant amount of He in the envelope. Comparing observations of NS--WD binaries could help to infer whether or not a binary underwent a ULX phase in the past.
   \item The tidal coupling between the spin and the orbit has a second-order effect on the stability of the binary. Our models contradict some previous results in the literature.
\end{itemize}

\begin{acknowledgements}
The authors thank the anonymous referee for their constructive comments that helped improve the manuscript. The authors are also grateful to Pablo Marchant, Vasilopoulos Georgios, and Andrew King for their useful discussions. This work was supported by the Swiss National Science Foundation Professorship grant (project number PP00P2 176868). This project has received funding from the European Union’s Horizon 2020 research and innovation program under the Marie Sklodowska-Curie RISE action, grant agreement no. 691164 (ASTROSTAT). The authors acknowledge the International Space Science Institute (ISSI) for supporting and funding the international teams program "Ultraluminous X-ray Sources: From the Local Group to the Very First Galaxies".
D.M. thanks the LSSTC Data Science Fellowship Program, which is funded by LSSTC, NSF Cybertraining Grant \#1829740, the Brinson Foundation, and the Moore Foundation; her participation in the program has benefited this work.
T.M.T. acknowledges an AIAS-COFUND Senior Fellowship funded by the European Union’s Horizon 2020 Research and Innovation Programme (grant no. 754513) and the Aarhus University Research Foundation. E.Z. acknowledges support from the Federal Commission for Scholarships for Foreign Students for the Swiss Government Excellence Scholarship (ESKAS No. 2019.0091) for the academic year 2019-2020.

\end{acknowledgements}

\bibliographystyle{aa}
\bibliography{aic}

\begin{thebibliography}{110}
\expandafter\ifx\csname natexlab\endcsname\relax\def\natexlab#1{#1}\fi

\bibitem[{{Abbott} {et~al.}(2016){Abbott}, {Abbott}, {Abbott}, {Abernathy},
  {Acernese}, {Ackley}, {Adams}, {Adams}, {Addesso}, {Adhikari}, \&
  et~al.}]{2016PhRvD..93l2003A}
{Abbott}, B.~P., {Abbott}, R., {Abbott}, T.~D., {et~al.} 2016, \prd, 93, 122003

\bibitem[{{Andrews} {et~al.}(2019){Andrews}, {Breivik}, \&
  {Chatterjee}}]{2019ApJ...886...68A}
{Andrews}, J.~J., {Breivik}, K., \& {Chatterjee}, S. 2019, \apj, 886, 68

\bibitem[{{Antoniadis} {et~al.}(2013){Antoniadis}, {Freire}, {Wex}, {Tauris},
  {Lynch}, {van Kerkwijk}, {Kramer}, {Bassa}, {Dhillon}, {Driebe}, {Hessels},
  {Kaspi}, {Kondratiev}, {Langer}, {Marsh}, {McLaughlin}, {Pennucci}, {Ransom},
  {Stairs}, {van Leeuwen}, {Verbiest}, \& {Whelan}}]{2013Sci...340..448A}
{Antoniadis}, J., {Freire}, P. C.~C., {Wex}, N., {et~al.} 2013, Science, 340,
  448

\bibitem[{{Bachetti} {et~al.}(2014){Bachetti}, {Harrison}, {Walton},
  {Grefenstette}, {Chakrabarty}, {F{\"u}rst}, {Barret}, {Beloborodov}, {Boggs},
  {Christensen}, {Craig}, {Fabian}, {Hailey}, {Hornschemeier}, {Kaspi},
  {Kulkarni}, {Maccarone}, {Miller}, {Rana}, {Stern}, {Tendulkar}, {Tomsick},
  {Webb}, \& {Zhang}}]{2014Natur.514..202B}
{Bachetti}, M., {Harrison}, F.~A., {Walton}, D.~J., {et~al.} 2014, \nat, 514,
  202

\bibitem[{{Bachetti} {et~al.}(2019){Bachetti}, {Maccarone}, {Brightman},
  {Brumback}, {F{\"u}rst}, {Harrison}, {Heida}, {Israel}, {Middleton},
  {Tomsick}, {Webb}, \& {Walton}}]{2019arXiv190506423B}
{Bachetti}, M., {Maccarone}, T.~J., {Brightman}, M., {et~al.} 2019, arXiv
  e-prints, arXiv:1905.06423

\bibitem[{{Basko} \& {Sunyaev}(1976)}]{1976MNRAS.175..395B}
{Basko}, M.~M. \& {Sunyaev}, R.~A. 1976, \mnras, 175, 395

\bibitem[{{Begelman}(2002)}]{2002ApJ...568L..97B}
{Begelman}, M.~C. 2002, \apjl, 568, L97

\bibitem[{{Begelman} {et~al.}(2006){Begelman}, {King}, \&
  {Pringle}}]{2006MNRAS.370..399B}
{Begelman}, M.~C., {King}, A.~R., \& {Pringle}, J.~E. 2006, \mnras, 370, 399

\bibitem[{{Brightman} {et~al.}(2018){Brightman}, {Harrison}, {F{\"u}rst},
  {Middleton}, {Walton}, {Stern}, {Fabian}, {Heida}, {Barret}, \&
  {Bachetti}}]{2018NatAs...2..312B}
{Brightman}, M., {Harrison}, F.~A., {F{\"u}rst}, F., {et~al.} 2018, Nature
  Astronomy, 2, 312

\bibitem[{{Carpano} {et~al.}(2018){Carpano}, {Haberl}, {Maitra}, \&
  {Vasilopoulos}}]{2018MNRAS.476L..45C}
{Carpano}, S., {Haberl}, F., {Maitra}, C., \& {Vasilopoulos}, G. 2018, \mnras,
  476, L45

\bibitem[{{Casares} {et~al.}(1998){Casares}, {Charles}, \&
  {Kuulkers}}]{1998ApJ...493L..39C}
{Casares}, J., {Charles}, P., \& {Kuulkers}, E. 1998, \apjl, 493, L39

\bibitem[{{Chashkina} {et~al.}(2019){Chashkina}, {Lipunova}, {Abolmasov}, \&
  {Poutanen}}]{2019A&A...626A..18C}
{Chashkina}, A., {Lipunova}, G., {Abolmasov}, P., \& {Poutanen}, J. 2019, \aap,
  626, A18

\bibitem[{{Choi} {et~al.}(2016){Choi}, {Dotter}, {Conroy}, {Cantiello},
  {Paxton}, \& {Johnson}}]{2016ApJ...823..102C}
{Choi}, J., {Dotter}, A., {Conroy}, C., {et~al.} 2016, \apj, 823, 102

\bibitem[{{Christodoulou} {et~al.}(2014){Christodoulou}, {Laycock}, \&
  {Kazanas}}]{2014arXiv1411.5434C}
{Christodoulou}, D.~M., {Laycock}, S. G.~T., \& {Kazanas}, D. 2014, arXiv
  e-prints, arXiv:1411.5434

\bibitem[{{Colbert} \& {Mushotzky}(1999)}]{1999ApJ...519...89C}
{Colbert}, E. J.~M. \& {Mushotzky}, R.~F. 1999, \apj, 519, 89

\bibitem[{{Cromartie} {et~al.}(2019){Cromartie}, {Fonseca}, {Ransom},
  {Demorest}, {Arzoumanian}, {Blumer}, {Brook}, {DeCesar}, {Dolch}, {Ellis},
  {Ferdman}, {Ferrara}, {Garver-Daniels}, {Gentile}, {Jones}, {Lam}, {Lorimer},
  {Lynch}, {McLaughlin}, {Ng}, {Nice}, {Pennucci}, {Spiewak}, {Stairs},
  {Stovall}, {Swiggum}, \& {Zhu}}]{2019arXiv190406759C}
{Cromartie}, H.~T., {Fonseca}, E., {Ransom}, S.~M., {et~al.} 2019, arXiv
  e-prints, arXiv:1904.06759

\bibitem[{{Doroshenko} {et~al.}(2018){Doroshenko}, {Tsygankov}, \&
  {Santangelo}}]{2018A&A...613A..19D}
{Doroshenko}, V., {Tsygankov}, S., \& {Santangelo}, A. 2018, \aap, 613, A19

\bibitem[{{Dotter}(2016)}]{2016ApJS..222....8D}
{Dotter}, A. 2016, \apjs, 222, 8

\bibitem[{{Ebisuzaki} {et~al.}(2001){Ebisuzaki}, {Makino}, {Tsuru}, {Funato},
  {Portegies Zwart}, {Hut}, {McMillan}, {Matsushita}, {Matsumoto}, \&
  {Kawabe}}]{2001ApJ...562L..19E}
{Ebisuzaki}, T., {Makino}, J., {Tsuru}, T.~G., {et~al.} 2001, \apjl, 562, L19

\bibitem[{{Eggleton}(1971)}]{1971MNRAS.151..351E}
{Eggleton}, P.~P. 1971, \mnras, 151, 351

\bibitem[{{Eggleton}(1972)}]{1972MNRAS.156..361E}
{Eggleton}, P.~P. 1972, \mnras, 156, 361

\bibitem[{{Eggleton}(1983)}]{1983ApJ...268..368E}
{Eggleton}, P.~P. 1983, \apj, 268, 368

\bibitem[{{Fabbiano}(1989)}]{1989ARA&A..27...87F}
{Fabbiano}, G. 1989, \araa, 27, 87

\bibitem[{{Finke} \& {Razzaque}(2017)}]{2017MNRAS.472.3683F}
{Finke}, J.~D. \& {Razzaque}, S. 2017, \mnras, 472, 3683

\bibitem[{{Fragos} {et~al.}(2019){Fragos}, {Andrews}, {Ramirez-Ruiz}, {Meynet},
  {Kalogera}, {Taam}, \& {Zezas}}]{2019ApJ...883L..45F}
{Fragos}, T., {Andrews}, J.~J., {Ramirez-Ruiz}, E., {et~al.} 2019, \apjl, 883,
  L45

\bibitem[{{Fragos} {et~al.}(2015){Fragos}, {Linden}, {Kalogera}, \&
  {Sklias}}]{2015ApJ...802L...5F}
{Fragos}, T., {Linden}, T., {Kalogera}, V., \& {Sklias}, P. 2015, \apjl, 802,
  L5

\bibitem[{{F{\"u}rst} {et~al.}(2016){F{\"u}rst}, {Walton}, {Harrison}, {Stern},
  {Barret}, {Brightman}, {Fabian}, {Grefenstette}, {Madsen}, {Middleton},
  {Miller}, {Pottschmidt}, {Ptak}, {Rana}, \& {Webb}}]{2016ApJ...831L..14F}
{F{\"u}rst}, F., {Walton}, D.~J., {Harrison}, F.~A., {et~al.} 2016, \apjl, 831,
  L14

\bibitem[{{Ge} {et~al.}(2020){Ge}, {Webbink}, \& {Han}}]{2020arXiv200600774G}
{Ge}, H., {Webbink}, R.~F., \& {Han}, Z. 2020, arXiv e-prints, arXiv:2006.00774

\bibitem[{{Ge} {et~al.}(2017){Ge}, {Zhang}, {Lu}, {Zhang}, {Weng}, {Xiong},
  {Liu}, {Song}, \& {HXMT-Collaboration}}]{2017ATel10907....1G}
{Ge}, M., {Zhang}, S., {Lu}, F., {et~al.} 2017, The Astronomer's Telegram,
  10907, 1

\bibitem[{{Gris{\'e}} {et~al.}(2008){Gris{\'e}}, {Pakull}, {Soria}, {Motch},
  {Smith}, {Ryder}, \& {B{\"o}ttcher}}]{2008A&A...486..151G}
{Gris{\'e}}, F., {Pakull}, M.~W., {Soria}, R., {et~al.} 2008, \aap, 486, 151

\bibitem[{{Heida} {et~al.}(2019){Heida}, {Lau}, {Davies}, {Brightman},
  {F{\"u}rst}, {Grefenstette}, {Kennea}, {Tramper}, {Walton}, \&
  {Harrison}}]{2019arXiv190902171H}
{Heida}, M., {Lau}, R.~M., {Davies}, B., {et~al.} 2019, arXiv e-prints,
  arXiv:1909.02171

\bibitem[{{Heil} {et~al.}(2009){Heil}, {Vaughan}, \&
  {Roberts}}]{2009MNRAS.397.1061H}
{Heil}, L.~M., {Vaughan}, S., \& {Roberts}, T.~P. 2009, \mnras, 397, 1061

\bibitem[{{Heusgen}(2016)}]{heu16}
{Heusgen}, F.~A. 2016, {Evolution of Low-Mass Helium Stars}

\bibitem[{{Huang} \& {Zeng}(2000)}]{2000ScChA..43..331H}
{Huang}, R. \& {Zeng}, Y. 2000, Science in China A: Mathematics, 43, 331

\bibitem[{{Hurley} {et~al.}(2002){Hurley}, {Tout}, \&
  {Pols}}]{2002MNRAS.329..897H}
{Hurley}, J.~R., {Tout}, C.~A., \& {Pols}, O.~R. 2002, \mnras, 329, 897

\bibitem[{{Hut}(1981)}]{1981A&A....99..126H}
{Hut}, P. 1981, \aap, 99, 126

\bibitem[{{Israel} {et~al.}(2017{\natexlab{a}}){Israel}, {Belfiore}, {Stella},
  {Esposito}, {Casella}, {De Luca}, {Marelli}, {Papitto}, {Perri}, {Puccetti},
  {Castillo}, {Salvetti}, {Tiengo}, {Zampieri}, {D'Agostino}, {Greiner},
  {Haberl}, {Novara}, {Salvaterra}, {Turolla}, {Watson}, {Wilms}, \&
  {Wolter}}]{2017Sci...355..817I}
{Israel}, G.~L., {Belfiore}, A., {Stella}, L., {et~al.} 2017{\natexlab{a}},
  Science, 355, 817

\bibitem[{{Israel} {et~al.}(2017{\natexlab{b}}){Israel}, {Papitto}, {Esposito},
  {Stella}, {Zampieri}, {Belfiore}, {Rodr{\'\i}guez Castillo}, {De Luca},
  {Tiengo}, {Haberl}, {Greiner}, {Salvaterra}, {Sandrelli}, \&
  {Lisini}}]{2017MNRAS.466L..48I}
{Israel}, G.~L., {Papitto}, A., {Esposito}, P., {et~al.} 2017{\natexlab{b}},
  \mnras, 466, L48

\bibitem[{{Istrate} {et~al.}(2014){Istrate}, {Tauris}, \&
  {Langer}}]{2014A&A...571A..45I}
{Istrate}, A.~G., {Tauris}, T.~M., \& {Langer}, N. 2014, \aap, 571, A45

\bibitem[{{Jenke} \& {Wilson-Hodge}(2017)}]{2017ATel10812....1J}
{Jenke}, P. \& {Wilson-Hodge}, C.~A. 2017, The Astronomer's Telegram, 10812, 1

\bibitem[{{Kaaret} {et~al.}(2017){Kaaret}, {Feng}, \&
  {Roberts}}]{2017ARA&A..55..303K}
{Kaaret}, P., {Feng}, H., \& {Roberts}, T.~P. 2017, \araa, 55, 303

\bibitem[{{Kennea} {et~al.}(2017){Kennea}, {Lien}, {Krimm}, {Cenko}, \&
  {Siegel}}]{2017ATel10809....1K}
{Kennea}, J.~A., {Lien}, A.~Y., {Krimm}, H.~A., {Cenko}, S.~B., \& {Siegel},
  M.~H. 2017, The Astronomer's Telegram, 10809, 1

\bibitem[{{King} \& {Lasota}(2019)}]{2019MNRAS.485.3588K}
{King}, A. \& {Lasota}, J.-P. 2019, \mnras, 485, 3588

\bibitem[{{King} {et~al.}(2017){King}, {Lasota}, \&
  {Klu{\'z}niak}}]{2017MNRAS.468L..59K}
{King}, A., {Lasota}, J.-P., \& {Klu{\'z}niak}, W. 2017, \mnras, 468, L59

\bibitem[{{King}(2009)}]{2009MNRAS.393L..41K}
{King}, A.~R. 2009, \mnras, 393, L41

\bibitem[{{King} \& {Begelman}(1999)}]{1999ApJ...519L.169K}
{King}, A.~R. \& {Begelman}, M.~C. 1999, \apjl, 519, L169

\bibitem[{{King} {et~al.}(2001){King}, {Davies}, {Ward}, {Fabbiano}, \&
  {Elvis}}]{2001ApJ...552L.109K}
{King}, A.~R., {Davies}, M.~B., {Ward}, M.~J., {Fabbiano}, G., \& {Elvis}, M.
  2001, \apjl, 552, L109

\bibitem[{{King} \& {Ritter}(1999)}]{1999MNRAS.309..253K}
{King}, A.~R. \& {Ritter}, H. 1999, \mnras, 309, 253

\bibitem[{{Kolb} {et~al.}(2000){Kolb}, {Davies}, {King}, \&
  {Ritter}}]{2000MNRAS.317..438K}
{Kolb}, U., {Davies}, M.~B., {King}, A., \& {Ritter}, H. 2000, \mnras, 317, 438

\bibitem[{{Kolb} \& {Ritter}(1990)}]{1990A&A...236..385K}
{Kolb}, U. \& {Ritter}, H. 1990, \aap, 236, 385

\bibitem[{{Kruckow} {et~al.}(2018){Kruckow}, {Tauris}, {Langer}, {Kramer}, \&
  {Izzard}}]{2018MNRAS.481.1908K}
{Kruckow}, M.~U., {Tauris}, T.~M., {Langer}, N., {Kramer}, M., \& {Izzard},
  R.~G. 2018, \mnras, 481, 1908

\bibitem[{{Lattimer} \& {Prakash}(2010)}]{2010arXiv1012.3208L}
{Lattimer}, J.~M. \& {Prakash}, M. 2010, arXiv e-prints, arXiv:1012.3208

\bibitem[{{Ledoux}(1947)}]{1947ApJ...105..305L}
{Ledoux}, P. 1947, \apj, 105, 305

\bibitem[{{Lyutikov}(2014)}]{2014arXiv1410.8745L}
{Lyutikov}, M. 2014, arXiv e-prints, arXiv:1410.8745

\bibitem[{{Maccarone} {et~al.}(2007){Maccarone}, {Kundu}, {Zepf}, \&
  {Rhode}}]{2007Natur.445..183M}
{Maccarone}, T.~J., {Kundu}, A., {Zepf}, S.~E., \& {Rhode}, K.~L. 2007, \nat,
  445, 183

\bibitem[{{Mandel} \& {de Mink}(2016)}]{2016MNRAS.458.2634M}
{Mandel}, I. \& {de Mink}, S.~E. 2016, \mnras, 458, 2634

\bibitem[{{Marchant} {et~al.}(2017){Marchant}, {Langer}, {Podsiadlowski},
  {Tauris}, {de Mink}, {Mandel}, \& {Moriya}}]{2017A&A...604A..55M}
{Marchant}, P., {Langer}, N., {Podsiadlowski}, P., {et~al.} 2017, \aap, 604,
  A55

\bibitem[{{Marchant} {et~al.}(2016){Marchant}, {Langer}, {Podsiadlowski},
  {Tauris}, \& {Moriya}}]{2016A&A...588A..50M}
{Marchant}, P., {Langer}, N., {Podsiadlowski}, P., {Tauris}, T.~M., \&
  {Moriya}, T.~J. 2016, \aap, 588, A50

\bibitem[{{Margalit} \& {Metzger}(2017)}]{2017ApJ...850L..19M}
{Margalit}, B. \& {Metzger}, B.~D. 2017, \apjl, 850, L19

\bibitem[{{Martinez} {et~al.}(2015){Martinez}, {Stovall}, {Freire}, {Deneva},
  {Jenet}, {McLaughlin}, {Bagchi}, {Bates}, \& {Ridolfi}}]{2015ApJ...812..143M}
{Martinez}, J.~G., {Stovall}, K., {Freire}, P.~C.~C., {et~al.} 2015, \apj, 812,
  143

\bibitem[{{Miller}(2006)}]{2006ASPC..352..121M}
{Miller}, J.~M. 2006, in Astronomical Society of the Pacific Conference Series,
  Vol. 352, New Horizons in Astronomy: Frank N. Bash Symposium, ed. S.~J.
  {Kannappan}, S.~{Redfield}, J.~E. {Kessler-Silacci}, M.~{Landriau}, \&
  N.~{Drory}, 121

\bibitem[{{Motch} {et~al.}(2011){Motch}, {Pakull}, {Gris{\'e}}, \&
  {Soria}}]{2011AN....332..367M}
{Motch}, C., {Pakull}, M.~W., {Gris{\'e}}, F., \& {Soria}, R. 2011,
  Astronomische Nachrichten, 332, 367

\bibitem[{{Motch} {et~al.}(2014){Motch}, {Pakull}, {Soria}, {Gris{\'e}}, \&
  {Pietrzy{\'n}ski}}]{2014Natur.514..198M}
{Motch}, C., {Pakull}, M.~W., {Soria}, R., {Gris{\'e}}, F., \&
  {Pietrzy{\'n}ski}, G. 2014, \nat, 514, 198

\bibitem[{{Nugis} \& {Lamers}(2000)}]{2000A&A...360..227N}
{Nugis}, T. \& {Lamers}, H.~J.~G.~L.~M. 2000, \aap, 360, 227

\bibitem[{{Orosz} \& {Kuulkers}(1999)}]{1999MNRAS.305..132O}
{Orosz}, J.~A. \& {Kuulkers}, E. 1999, \mnras, 305, 132

\bibitem[{{Pakull} \& {Gris{\'e}}(2008)}]{2008AIPC.1010..303P}
{Pakull}, M.~W. \& {Gris{\'e}}, F. 2008, in American Institute of Physics
  Conference Series, Vol. 1010, A Population Explosion: The Nature \& Evolution
  of X-ray Binaries in Diverse Environments, ed. R.~M. {Bandyopadhyay},
  S.~{Wachter}, D.~{Gelino}, \& C.~R. {Gelino}, 303--307

\bibitem[{{Paxton} {et~al.}(2011){Paxton}, {Bildsten}, {Dotter}, {Herwig},
  {Lesaffre}, \& {Timmes}}]{2011ApJS..192....3P}
{Paxton}, B., {Bildsten}, L., {Dotter}, A., {et~al.} 2011, \apjs, 192, 3

\bibitem[{{Paxton} {et~al.}(2013){Paxton}, {Cantiello}, {Arras}, {Bildsten},
  {Brown}, {Dotter}, {Mankovich}, {Montgomery}, {Stello}, {Timmes}, \&
  {Townsend}}]{2013ApJS..208....4P}
{Paxton}, B., {Cantiello}, M., {Arras}, P., {et~al.} 2013, \apjs, 208, 4

\bibitem[{{Paxton} {et~al.}(2015){Paxton}, {Marchant}, {Schwab}, {Bauer},
  {Bildsten}, {Cantiello}, {Dessart}, {Farmer}, {Hu}, {Langer}, {Townsend},
  {Townsley}, \& {Timmes}}]{2015ApJS..220...15P}
{Paxton}, B., {Marchant}, P., {Schwab}, J., {et~al.} 2015, \apjs, 220, 15

\bibitem[{{Paxton} {et~al.}(2018){Paxton}, {Schwab}, {Bauer}, {Bildsten},
  {Blinnikov}, {Duffell}, {Farmer}, {Goldberg}, {Marchant}, {Sorokina},
  {Thoul}, {Townsend}, \& {Timmes}}]{2018ApJS..234...34P}
{Paxton}, B., {Schwab}, J., {Bauer}, E.~B., {et~al.} 2018, \apjs, 234, 34

\bibitem[{{Paxton} {et~al.}(2019){Paxton}, {Smolec}, {Schwab}, {Gautschy},
  {Bildsten}, {Cantiello}, {Dotter}, {Farmer}, {Goldberg}, {Jermyn}, {Kanbur},
  {Marchant}, {Thoul}, {Townsend}, {Wolf}, {Zhang}, \&
  {Timmes}}]{2019ApJS..243...10P}
{Paxton}, B., {Smolec}, R., {Schwab}, J., {et~al.} 2019, \apjs, 243, 10

\bibitem[{{Podsiadlowski} \& {Rappaport}(2000)}]{2000ApJ...529..946P}
{Podsiadlowski}, P. \& {Rappaport}, S. 2000, \apj, 529, 946

\bibitem[{{Pols} {et~al.}(1998){Pols}, {Schr{\"o}der}, {Hurley}, {Tout}, \&
  {Eggleton}}]{1998MNRAS.298..525P}
{Pols}, O.~R., {Schr{\"o}der}, K.-P., {Hurley}, J.~R., {Tout}, C.~A., \&
  {Eggleton}, P.~P. 1998, \mnras, 298, 525

\bibitem[{{Poutanen} {et~al.}(2007){Poutanen}, {Lipunova}, {Fabrika},
  {Butkevich}, \& {Abolmasov}}]{2007MNRAS.377.1187P}
{Poutanen}, J., {Lipunova}, G., {Fabrika}, S., {Butkevich}, A.~G., \&
  {Abolmasov}, P. 2007, \mnras, 377, 1187

\bibitem[{{Pratt} \& {Strittmatter}(1976)}]{1976ApJ...204L..29P}
{Pratt}, J.~P. \& {Strittmatter}, P.~A. 1976, \apjl, 204, L29

\bibitem[{{Qin} {et~al.}(2018){Qin}, {Fragos}, {Meynet}, {Andrews},
  {S{\o}rensen}, \& {Song}}]{2018A&A...616A..28Q}
{Qin}, Y., {Fragos}, T., {Meynet}, G., {et~al.} 2018, \aap, 616, A28

\bibitem[{{Quast} {et~al.}(2019){Quast}, {langer}, \&
  {Tauris}}]{2019arXiv190304995Q}
{Quast}, M., {langer}, N., \& {Tauris}, T.~M. 2019, arXiv e-prints,
  arXiv:1903.04995

\bibitem[{{Rappaport} {et~al.}(1983){Rappaport}, {Verbunt}, \&
  {Joss}}]{1983ApJ...275..713R}
{Rappaport}, S., {Verbunt}, F., \& {Joss}, P.~C. 1983, \apj, 275, 713

\bibitem[{{Rasio} {et~al.}(1996){Rasio}, {Tout}, {Lubow}, \&
  {Livio}}]{1996ApJ...470.1187R}
{Rasio}, F.~A., {Tout}, C.~A., {Lubow}, S.~H., \& {Livio}, M. 1996, \apj, 470,
  1187

\bibitem[{{Ray} {et~al.}(2019){Ray}, {Guillot}, {Ho}, {Kerr}, {Enoto},
  {Gendreau}, {Arzoumanian}, {Altamirano}, {Bogdanov}, {Campion},
  {Chakrabarty}, {Deneva}, {Jaisawal}, {Kozon}, {Malacaria}, {Strohmayer}, \&
  {Wolff}}]{2019ApJ...879..130R}
{Ray}, P.~S., {Guillot}, S., {Ho}, W. C.~G., {et~al.} 2019, \apj, 879, 130

\bibitem[{{Reimers}(1975)}]{1975psae.book..229R}
{Reimers}, D. 1975, {Circumstellar envelopes and mass loss of red giant
  stars.}, 229--256

\bibitem[{{Rezzolla} {et~al.}(2018){Rezzolla}, {Most}, \&
  {Weih}}]{2018ApJ...852L..25R}
{Rezzolla}, L., {Most}, E.~R., \& {Weih}, L.~R. 2018, \apjl, 852, L25

\bibitem[{{Ritter} \& {King}(2001)}]{2001ASPC..229..423R}
{Ritter}, H. \& {King}, A.~R. 2001, in Astronomical Society of the Pacific
  Conference Series, Vol. 229, Evolution of Binary and Multiple Star Systems,
  ed. P.~{Podsiadlowski}, S.~{Rappaport}, A.~R. {King}, F.~{D'Antona}, \&
  L.~{Burderi}, 423

\bibitem[{{Rodr{\'\i}guez Castillo} {et~al.}(2019){Rodr{\'\i}guez Castillo},
  {Israel}, {Belfiore}, {Bernardini}, {Esposito}, {Pintore}, {De Luca},
  {Papitto}, {Stella}, {Tiengo}, {Zampieri}, {Bachetti}, {Brightman},
  {Casella}, {D'Agostino}, {Dall'Osso}, {Earnshaw}, {F{\"u}rst}, {Haberl},
  {Harrison}, {Mapelli}, {Marelli}, {Middleton}, {Pinto}, {Roberts},
  {Salvaterra}, {Turolla}, {Walton}, \& {Wolter}}]{2019arXiv190604791R}
{Rodr{\'\i}guez Castillo}, G.~A., {Israel}, G.~L., {Belfiore}, A., {et~al.}
  2019, arXiv e-prints, arXiv:1906.04791

\bibitem[{{Sathyaprakash} {et~al.}(2019){Sathyaprakash}, {Roberts}, {Walton},
  {Fuerst}, {Bachetti}, {Pinto}, {Alston}, {Earnshaw}, {Fabian}, {Middleton},
  \& {Soria}}]{2019MNRAS.tmpL.104S}
{Sathyaprakash}, R., {Roberts}, T.~P., {Walton}, D.~J., {et~al.} 2019, \mnras,
  L104

\bibitem[{{Shakura} \& {Sunyaev}(1973)}]{1973A&A....24..337S}
{Shakura}, N.~I. \& {Sunyaev}, R.~A. 1973, \aap, 24, 337

\bibitem[{{Shao} \& {Li}(2012)}]{2012ApJ...756...85S}
{Shao}, Y. \& {Li}, X.-D. 2012, \apj, 756, 85

\bibitem[{{Shao} \& {Li}(2015)}]{2015ApJ...802..131S}
{Shao}, Y. \& {Li}, X.-D. 2015, \apj, 802, 131

\bibitem[{{Sutton} {et~al.}(2013){Sutton}, {Roberts}, {Gladstone}, {Farrell},
  {Reilly}, {Goad}, \& {Gehrels}}]{2013MNRAS.434.1702S}
{Sutton}, A.~D., {Roberts}, T.~P., {Gladstone}, J.~C., {et~al.} 2013, \mnras,
  434, 1702

\bibitem[{{Takahashi} \& {Ohsuga}(2017)}]{2017ApJ...845L...9T}
{Takahashi}, H.~R. \& {Ohsuga}, K. 2017, \apjl, 845, L9

\bibitem[{{Tao} {et~al.}(2019){Tao}, {Feng}, {Zhang}, {Bu}, {Zhang}, {Qu}, \&
  {Zhang}}]{2019ApJ...873...19T}
{Tao}, L., {Feng}, H., {Zhang}, S., {et~al.} 2019, \apj, 873, 19

\bibitem[{{Tauris} \& {Janka}(2019)}]{2019ApJ...886L..20T}
{Tauris}, T.~M. \& {Janka}, H.-T. 2019, \apjl, 886, L20

\bibitem[{{Tauris} {et~al.}(2017){Tauris}, {Kramer}, {Freire}, {Wex}, {Janka},
  {Langer}, {Podsiadlowski}, {Bozzo}, {Chaty}, {Kruckow}, {van den Heuvel},
  {Antoniadis}, {Breton}, \& {Champion}}]{2017ApJ...846..170T}
{Tauris}, T.~M., {Kramer}, M., {Freire}, P.~C.~C., {et~al.} 2017, \apj, 846,
  170

\bibitem[{{Tauris} \& {Savonije}(1999)}]{1999A&A...350..928T}
{Tauris}, T.~M. \& {Savonije}, G.~J. 1999, \aap, 350, 928

\bibitem[{{Tauris} \& {Savonije}(2001)}]{2001nsbh.conf..337T}
{Tauris}, T.~M. \& {Savonije}, G.~J. 2001, in The Neutron Star - Black Hole
  Connection, ed. C.~{Kouveliotou}, J.~{Ventura}, \& E.~{van den Heuvel}, Vol.
  567, 337

\bibitem[{{Tauris} \& {van den Heuvel}(2006)}]{2006csxs.book..623T}
{Tauris}, T.~M. \& {van den Heuvel}, E.~P.~J. 2006, {Formation and evolution of
  compact stellar X-ray sources}, Vol.~39, 623--665

\bibitem[{{Tauris} {et~al.}(2000){Tauris}, {van den Heuvel}, \&
  {Savonije}}]{2000ApJ...530L..93T}
{Tauris}, T.~M., {van den Heuvel}, E. P.~J., \& {Savonije}, G.~J. 2000, \apjl,
  530, L93

\bibitem[{{Thorne} \& {Zytkow}(1977)}]{1977ApJ...212..832T}
{Thorne}, K.~S. \& {Zytkow}, A.~N. 1977, \apj, 212, 832

\bibitem[{{Vasilopoulos} {et~al.}(2018){Vasilopoulos}, {Haberl}, {Carpano}, \&
  {Maitra}}]{2018A&A...620L..12V}
{Vasilopoulos}, G., {Haberl}, F., {Carpano}, S., \& {Maitra}, C. 2018, \aap,
  620, L12

\bibitem[{{Vasilopoulos} {et~al.}(2020){Vasilopoulos}, {Lander}, {Koliopanos},
  \& {Bailyn}}]{2020MNRAS.491.4949V}
{Vasilopoulos}, G., {Lander}, S.~K., {Koliopanos}, F., \& {Bailyn}, C.~D. 2020,
  \mnras, 491, 4949

\bibitem[{{Vasilopoulos} {et~al.}(2019){Vasilopoulos}, {Petropoulou},
  {Koliopanos}, {Ray}, {Bailyn}, {Haberl}, \& {Gendreau}}]{2019MNRAS.488.5225V}
{Vasilopoulos}, G., {Petropoulou}, M., {Koliopanos}, F., {et~al.} 2019, \mnras,
  488, 5225

\bibitem[{{Verbunt} \& {Phinney}(1995)}]{1995A&A...296..709V}
{Verbunt}, F. \& {Phinney}, E.~S. 1995, \aap, 296, 709

\bibitem[{{Vinokurov} {et~al.}(2013){Vinokurov}, {Fabrika}, \&
  {Atapin}}]{2013AstBu..68..139V}
{Vinokurov}, A., {Fabrika}, S., \& {Atapin}, K. 2013, Astrophysical Bulletin,
  68, 139

\bibitem[{{Walton} {et~al.}(2012){Walton}, {Miller}, {Reis}, \&
  {Fabian}}]{2012MNRAS.426..473W}
{Walton}, D.~J., {Miller}, J.~M., {Reis}, R.~C., \& {Fabian}, A.~C. 2012,
  \mnras, 426, 473

\bibitem[{{Weng} \& {Zhang}(2011)}]{2011ApJ...739...42W}
{Weng}, S.-S. \& {Zhang}, S.-N. 2011, \apj, 739, 42

\bibitem[{{Wiktorowicz} {et~al.}(2019){Wiktorowicz}, {Lasota}, {Middleton}, \&
  {Belczynski}}]{2019ApJ...875...53W}
{Wiktorowicz}, G., {Lasota}, J.-P., {Middleton}, M., \& {Belczynski}, K. 2019,
  \apj, 875, 53

\bibitem[{{Wiktorowicz} {et~al.}(2017){Wiktorowicz}, {Sobolewska}, {Lasota}, \&
  {Belczynski}}]{2017ApJ...846...17W}
{Wiktorowicz}, G., {Sobolewska}, M., {Lasota}, J.-P., \& {Belczynski}, K. 2017,
  \apj, 846, 17

\bibitem[{{Wiktorowicz} {et~al.}(2015){Wiktorowicz}, {Sobolewska},
  {S{\k{a}}dowski}, \& {Belczynski}}]{2015ApJ...810...20W}
{Wiktorowicz}, G., {Sobolewska}, M., {S{\k{a}}dowski}, A., \& {Belczynski}, K.
  2015, \apj, 810, 20

\bibitem[{{Zahn}(1977)}]{1977A&A....57..383Z}
{Zahn}, J.~P. 1977, \aap, 500, 121

\bibitem[{{Zhang} {et~al.}(2019){Zhang}, {Ge}, {Song}, {Zhang}, {Qu}, {Zhang},
  {Doroshenko}, {Tao}, {Ji}, {G{\"u}ng{\"o}r}, {Santangelo}, {Shi}, {Chang},
  {Chen}, {Chen}, {Chen}, {Chen}, {Chen}, {Cui}, {Cui}, {Deng}, {Dong}, {Du},
  {Fu}, {Gao}, {Gao}, {Gao}, {Gu}, {Guan}, {Guo}, {Han}, {Hu}, {Huang}, {Huo},
  {Jia}, {Jiang}, {Jiang}, {Jin}, {Jin}, {Li}, {Li}, {Li}, {Li}, {Li}, {Li},
  {Li}, {Li}, {Li}, {Li}, {Li}, {Liang}, {Liao}, {Liu}, {Liu}, {Liu}, {Liu},
  {Liu}, {Liu}, {Liu}, {Lu}, {Lu}, {Luo}, {Ma}, {Meng}, {Nang}, {Nie}, {Ou},
  {Sai}, {Sun}, {Tan}, {Tao}, {Tuo}, {Wang}, {Wang}, {Wang}, {Wang}, {Wang},
  {Wen}, {Wu}, {Wu}, {Xiao}, {Xiong}, {Xu}, {Xu}, {Yan}, {Yang}, {Yang},
  {Yang}, {Zhang}, {Zhang}, {Zhang}, {Zhang}, {Zhang}, {Zhang}, {Zhang},
  {Zhang}, {Zhang}, {Zhang}, {Zhang}, {Zhang}, {Zhang}, {Zhang}, {Zhang},
  {Zhao}, {Zhao}, {Zhao}, {Zheng}, {Zhu}, {Zhu}, {Zou}, \& {(Insight-HXMT
  collaboration}}]{2019ApJ...879...61Z}
{Zhang}, Y., {Ge}, M., {Song}, L., {et~al.} 2019, \apj, 879, 61

\end{thebibliography}
\begin{appendix}
\section{Analytical solution of orbital evolution}
\label{append1}

The orbital angular momentum ($J_\mathrm{orb}$) of a binary with component masses as $M_{\mathrm{donor}}$ and $M_{\mathrm{acc}}$ can be described as,
\begin{equation}
J_\mathrm{orb}= M_{\mathrm{donor}} M_{\mathrm{acc}} \sqrt{\frac{Ga(1-e^2)}{M}}, \label{orb_am}
\end{equation}
where $M = M_{\mathrm{donor}} + M_{\mathrm{acc}}$, $a$ is the binary separation, and $e$ is the eccentricity of the orbit. For a circular orbit ($e=0$), a logarithmic differentiation of Eq.~(\ref{orb_am}) gives the rate of change of the orbital separation ($\dot{a}$),
\begin{equation}\label{adot}
    \frac{\dot{a}}{a}=2 \frac{\dot{J}_\mathrm{orb}}{J_\mathrm{orb}} - 2 \frac{\dot{M}_{\mathrm{donor}}}{M_{\mathrm{donor}}}-2\frac{\dot{M}_{\mathrm{acc}}}{M_{\mathrm{acc}}} + \frac{\dot{M}_{\mathrm{donor}}+\dot{M}_{\mathrm{acc}}}{M}.
\end{equation}

We consider a binary system which is in synchronous rotation with its components as tides are efficiently transferring angular momentum from donor to the orbit and vice versa. That is, it is tidally locked. Therefore, the orbital angular velocity ($\omega_{\rm orb}$) is equal to the spin angular velocity ($\omega_{\rm donor}$), which we take as $\omega_{\rm orb}=\omega_{\rm donor}=\omega$. 

The mass lost from the binary carries away angular momentum from the orbit. For a such a system containing a mass losing donor star ($M_{\rm donor}$) and an accreting neutron star (or any other compact object with mass, $M_{\rm acc}$), the changes in the orbital angular momentum are described by 
\begin{equation}\label{j_orb}
    \dot{J}_{\rm orb} = \omega a^{2}_{2}\dot{M}_{\rm donor}+ \dot{J}_{\rm ls},
\end{equation}
where $a_{2} = aM_{\rm donor}/M$ is the distance from where the mass transferred is lost from the system, that is from around $M_{\rm acc}$. 
The term $\dot{J}_{\rm ls}$ stands for the contribution to the rate of change of the total angular momentum ($\dot{J}_{\rm orb}$) from the donor spin ($\dot{J}_{\rm spin}$) and vice versa, due to the coupling between the spin and the orbit. Changes in the spin angular momentum of the donor are defined by
\begin{equation}\label{j_spin1}
    \dot{J}_{\rm spin} = - \dot{J}_{\rm ls}.
\end{equation}
The donor is undergoing solid-body rotation and so its angular momentum can be described as $J_{\rm spin}=I_{\rm donor}\omega$, and therefore $\dot{J}_{\rm spin}$ can also be expressed in general as
\begin{equation}\label{j_spin2}
    \dot{J}_{\rm spin}=\dot{I}_{\rm donor}\omega + I_{\rm donor}\dot{\omega},
\end{equation}
where $I_{\rm donor}$ and $\omega$ are the moment of inertia and angular velocity of the donor. Using Eqs.~(\ref{j_orb}) and (\ref{j_spin2}), the rate of change in total orbital angular momentum comes out to be
\begin{equation}\label{j_orb2}
    \dot{J}_{\rm orb} = \omega a^{2}_{2}\dot{M}_{\rm donor}-\dot{I}_{\rm donor}\omega - I_{\rm donor}\dot{\omega}.
\end{equation}
Substituting Eq.~(\ref{j_orb2}) into Eq.~(\ref{adot}) and using the assumption of fully non-conservative mass transfer, i.e. $\dot{M}_{\mathrm{acc}}=0$,
\begin{equation}
    \frac{\dot{a}}{a}=2 \frac{ \omega a^{2}_{2}\dot{M}_{\rm donor}-\dot{I}_{\rm donor}\omega - I_{\rm donor}\dot{\omega}}{J_\mathrm{orb}} - 2 \frac{\dot{M}_{\mathrm{donor}}}{M_{\mathrm{donor}}} + \frac{\dot{M}_{\mathrm{donor}}}{M}.
\end{equation}
Using the definitions of $J_{\rm orb}=M_{\mathrm{donor}} M_{\mathrm{acc}} \sqrt{Ga/M}$, $\omega=\sqrt{GM/a^3}$, the logarithmic differentiation of $\omega$, $\dot{\omega}=\omega(\dot{M}_{\rm donor}/M-3\dot{a}/a)/2$, and simplifying further we get the analytical solution for the orbital separation,
\begin{align}
    \nonumber \frac{\dot{a}}{a} = \bigg[\frac{M_{\mathrm{donor}} M_{\mathrm{acc}} a^{2}}{M_{\mathrm{donor}} M_{\mathrm{acc}} a^{2} - 3 I_{\mathrm{donor}} M}\bigg]\times \bigg[\dot{M}_{\mathrm{donor}} \bigg(\frac{2M_{\mathrm{donor}}}{M_{\mathrm{acc}}M} - \\ \frac{2}{M_{\mathrm{donor}}} +
    \frac{1}{M}\bigg) - \bigg(\frac{2\dot{I}_\mathrm{donor}M + I_\mathrm{donor}\dot{M}_{\mathrm{donor}}}{M_{\mathrm{donor}} M_{\mathrm{acc}} a^{2}}\bigg)\bigg].
\end{align}
This equation describes the evolution of orbital separation in a binary where there is efficient exchange of angular momentum between the donor and the orbit and the donor is undergoing solid-body rotation.
\subsection{Point mass approximation}
In case of no exchange of angular momentum between spin and orbit, the binary components are treated as point masses, therefore $I_{\rm donor}=0$. Using this assumption, the orbital evolution equation simplifies as follows,
\begin{align}\label{point_1}
    \frac{\dot{a}}{a} = \dot{M}_{\mathrm{donor}} \bigg(\frac{2M_{\mathrm{donor}}}{M_{\mathrm{acc}}M} - \frac{2}{M_{\mathrm{donor}}} +
    \frac{1}{M}\bigg).
\end{align}

If we look at Eq.~(16.18) in \citep{2006csxs.book..623T} which is,
\begin{equation}\label{tauris_jorb}
   \frac{\dot{J}_{\rm orb}}{J_{\rm orb}}=\bigg[\frac{\alpha + \beta (M_{\rm donor}/M_{\rm acc})^2 + \delta\gamma(1+(M_{\rm donor}/M_{\rm acc})^2)}{1+M_{\rm donor}/M_{\rm acc}}\bigg]\frac{\dot{M}_{\rm donor}}{M_{\rm donor}},
\end{equation}
where $\alpha$ is the fractional mass lost directly from the donor, $\beta$ is the fractional mass lost from the vicinity of the accretor, $\delta$ is the fractional mass lost from a circumbinary toroid, and the radius of the circumbinary toroid is defined by $\gamma^2 a$.

Using the assumptions $\dot{M}_{\mathrm{acc}}=0\implies \beta=1$, and $\alpha=\delta=0$, we substitute Eq.~(\ref{tauris_jorb}) in Eq.~(\ref{adot}) and simplify to get the following,
\begin{align}\label{point_2}
    \frac{\dot{a}}{a} = \dot{M}_{\mathrm{donor}} \bigg(\frac{2M_{\mathrm{donor}}}{M_{\mathrm{acc}}M} - \frac{2}{M_{\mathrm{donor}}} +
    \frac{1}{M}\bigg).
\end{align}
This equation is consistent with our calculations where we obtain Eq.~(\ref{point_1}).

\end{appendix}
\end{document}